# Baryon Octet to Decuplet Electromagnetic Transitions

DEREK B. LEINWEBER
*Department of Physics and Center for Theoretical Physics*
*University of Maryland, College Park, MD 20742*

TERRENCE DRAPER
*Department of Physics and Astronomy, University of Kentucky, Lexington, KY 40506*

R.M. WOLOSHYN
*TRIUMF, 4004 Wesbrook Mall, Vancouver, B.C. Canada V6T 2A3*

November, 1992

## Abstract

The electromagnetic transition moments of the $SU(3)$-flavor baryon octet to decuplet are examined within a lattice simulation of quenched QCD. The magnetic transition moment for the $N\gamma \leftrightarrow \Delta$ channel is found to be in agreement with recent experimental analyses. The lattice results indicate $\mu_{p\Delta}/\mu_p = 0.88(15)$. In terms of the Particle Data Group convention, $f_{M1} = 0.231(41)$ GeV$^{-1/2}$ for $p\gamma \leftrightarrow \Delta^+$ transitions. Lattice predictions for the hyperon $M1$ transition moments agree with those of a simple quark model. However the manner in which the quarks contribute to the transition moments in the lattice simulation is different from that anticipated by quark model calculations. The scalar quadrupole form factor exhibits a behavior consistent with previous multipole analyses. The $E2/M1$ multipole transition moment ratios are also determined. The lattice results suggest $R_{EM} \equiv -\mathcal{G}_{E2}/\mathcal{G}_{M1} = +3 \pm 8$ % for $p\gamma \leftrightarrow \Delta^+$ transitions. Of particular interest are significant nonvanishing signals for the $E2/M1$ ratio in $\Xi^-$ and $\Sigma^-$ electromagnetic transitions.

# I. INTRODUCTION

One of the great promises of the lattice gauge approach to QCD is to reveal the quark substructure and dynamics of hadrons. Lattice calculations of SU(3)-flavor octet and decuplet baryon electromagnetic form factors [1,2] have made significant steps in this direction. Calculations of quark charge distribution radii have described the manner in which the quarks are distributed within baryons and how these distributions change from one baryon to the next. The magnetic properties of quarks within baryons have also been examined. A strong sensitivity to the environment in which a quark resides is seen in the quark contributions to the octet baryon magnetic moments. Evidence suggests dynamical quark mass effects, nonperturbative gluon interactions and relativistic dynamics are the mechanisms underlying this phenomena. The lattice calculations have also given us access to many QCD observables that otherwise are not available at present from laboratory experiments. This new information will be indispensable in both the development and testing of model hypotheses for low-energy hadron physics.

In this paper we continue our exploration of baryon electromagnetic structure through calculations of the electromagnetic multipole form factors describing the $N\ \gamma \leftrightarrow \Delta$ transition as well as other octet to decuplet baryon electromagnetic transitions. The electromagnetic transitions of $N\ \gamma \leftrightarrow \Delta$ have been the subject of intense study since the preliminary analyses of the photoproduction data [3–5] in which a nonvanishing value for the $E2/M1$ ratio of form factors was found. The finite value of this ratio indicates some deviation from spherical symmetry in the nucleon and/or $\Delta$ ground state wave functions.

The calculation and prediction of transition moments is an integral part of the development and testing of model hypotheses. By calculating the transition moments of the entire baryon octet we hope to discover the dependence of the transition moments on the quark mass and make more extensive comparisons with model calculations. It will be interesting to learn whether the quark electromagnetic properties resemble those of quarks in octet baryons or decuplet baryons. A comparison of the effective quark moments determined from the transition moments with those determined in our previous analyses will give a great deal of insight into the concept of constituent quarks and intrinsic moments. Moreover, the lattice results will provide access to many more transition moments which may be useful for model development.

While most of the present attention is directed to a determination of the $E2/M1$ transition moment ratio it should be noted that the actual value of the $M1$ transition moment is itself not well known. Factors contributing to this are the long standing and well known problems associated with defining the properties of a hadron unstable to strong interactions. In this lattice calculation, these issues do not present a problem until we attempt to make a comparison with the experimental multipole analyses. For the quark masses considered on the lattice the $\Delta$ is stable, since energy conservation prevents it from decaying to a more massive $\pi N$ state.

An obvious approach to the definition of $\Delta$ properties on the lattice is to calculate where the $\Delta$ is stable and smoothly extrapolate the values to the physical quark masses. In essence, this defines a model for what we mean by the properties of the $\Delta$. Such a definition is common to many models of hadron structure, such as the simple quark model, which have excitations which are stable to strong interactions. However, the $\Delta$ should be regarded as dressed as the possibility of Z-graphs in the quark propagators allow intermediate states of the $\Delta$ with multiple $q\overline{q}$ states. These intermediate states are expected to have overlap with mesonic dressings of $\Delta$ including virtual $\pi N$ intermediate states [6]. The quark mass dependence of the $M1$ transition moment is extremely weak. Any problems associated with the extrapolation to the physical quark masses are not apparent.

Early lattice calculations of electromagnetic form factors focused on the pion with $SU(2)$ color [7–9] and later with $SU(3)$ color [10–12]. Calculations of the proton electric form factor followed [13]. Electromagnetic form factors of $\pi$, $\rho$ and $N$ were calculated [14] from which magnetic moments and electric charge radii were extracted. Our analysis of the entire baryon flavor-octet followed [1] in which electromagnetic properties were reported for both baryons and the quark sector contributions. The $q^2$ dependence of the nucleon electromagnetic form factors was examined using a method which characterizes one of the nucleon interpolating fields as a zero momentum secondary source [15]. Finally, our examination of the baryon decuplet [2] revealed underlying quark interactions in which mass effects and spin-dependent forces counteract each other. The significant baryon dependence of quark electromagnetic properties seen in the octet was not reproduced in decuplet baryons.

The study of hadronic wave functions has also received some attention and has confirmed the presence of spin-dependent interactions in lattice simulations of QCD that cause the distribution of the $d$ quarks in the neutron to be larger than that of the $u$ quark [16]. Similar spin-dependent interactions were seen in our calculations of octet baryon electromagnetic form factors and are responsible for the negative squared charge radius of the neutron. The wave function analyses also indicate an absence of substantial scalar diquark clustering, in agreement with our calculations of octet and decuplet baryon charge radii [17]. A comparison of wave functions calculated in quenched and full QCD [18] indicates there is little difference between the two calculations outside of a simple renormalization of the coupling constant. This supports the usefulness of the quenched approximation, at least for the quark masses currently investigated on the lattice.

The qualitative behavior of charge distributions in the nucleon have been confirmed in wave function analyses [16] which employ a very different approach in obtaining information on quark distributions. Unfortunately a quantitative comparison is not possible as wave functions are not defined in a gauge invariant manner. Wave functions calculated in different gauges have different shapes and distribution sizes. In contrast to our lattice form factor analyses, it is not possible to calculate, for example, a charge radius from the wave functions that may be directly compared with experimental measurements.

The format of our paper is as follows. In section II, we review the interpolating fields used to excite the octet and decuplet baryons. The two- and three-point correlation functions corresponding to these interpolating fields are presented at the quark level. The transition moments for $\Lambda\,\gamma \leftrightarrow \Sigma^{*0}$ are not reported here as the correlation functions for this transition differ from those of the rest of the octet. These transition moments will be examined in a future lattice investigation of $\Lambda_H$ baryons and heavy quark symmetry. Section III reviews the covariant matrix element that defines the baryon multipole transition moments. The formalism developed to isolate and extract the form factors from the current matrix element is presented in detail. Lattice techniques are briefly summarized at the end of section III. Calculations of the three-point correlation functions used to extract the form factors from the lattice simulations are illustrated in section IV. To provide some background to the study of transition moments, we have generalized the quark model of Darewych, Horbatsch and Koniuk [19] for $\Sigma$ and $\Lambda$ transitions, to include nucleon and $\Xi$ transitions. These quark model predictions are summarized in section V. Section VI presents the lattice determinations of the transition moments and compares the lattice results with model expectations. Section VII presents an overview of the results and an outlook on future calculations.

## II. CORRELATION FUNCTIONS AT THE QUARK LEVEL

### A. Interpolating fields

For the octet baryon interpolating fields we use the following standard forms. For the proton,

$$\chi^p(x) = \epsilon^{abc} \left( u^{Ta}(x) C \gamma_5 d^b(x) \right) u^c(x). \tag{2.1}$$

This interpolating field has the advantage of excluding components which vanish in a non-relativistic reduction and which otherwise act to increase statistical uncertainties in lattice simulations [1]. Unless otherwise noted, we follow the notation of Sakurai [20]. The Dirac gamma matrices are Hermitian and satisfy $\{\gamma_\mu, \gamma_\nu\} = 2\delta_{\mu\nu}$, with $\sigma_{\mu\nu} = \frac{1}{2i}[\gamma_\mu, \gamma_\nu]$. $C = \gamma_4 \gamma_2$ is the charge conjugation matrix, $a$, $b$, $c$ are color indices, $u(x)$ is a $u$-quark field, and the superscript $T$ denotes transpose. As in our decuplet baryon analysis, we utilize the following $\Delta^+$ interpolating field

$$\chi_\sigma^{\Delta^+}(x) = \frac{1}{\sqrt{3}} \epsilon^{abc} \Big[ 2 \left( u^{Ta}(x) C \gamma_\sigma d^b(x) \right) u^c(x) \\ + \left( u^{Ta}(x) C \gamma_\sigma u^b(x) \right) d^c(x) \Big]. \tag{2.2}$$

Other baryon interpolating fields are obtained with the appropriate substitutions of $u(x)$ or $d(x) \to u(x)$, $d(x)$ or $s(x)$. For the transition moments of $\Sigma^0 \gamma \leftrightarrow \Sigma^{*0}$ we use the octet interpolating field

$$\chi^{\Sigma^0}(x) = \sqrt{\frac{1}{2}} \epsilon^{abc} \Big[ \left( u^{Ta}(x) C \gamma_5 s^b(x) \right) d^c(x) \\ + \left( d^{Ta}(x) C \gamma_5 s^b(x) \right) u^c(x) \Big], \tag{2.3}$$

and the decuplet interpolating field

$$\chi_\sigma^{\Sigma^{*0}}(x) = \sqrt{\frac{2}{3}} \epsilon^{abc} \Big[ \left( u^{Ta}(x) C \gamma_\sigma d^b(x) \right) s^c(x) \\ + \left( d^{Ta}(x) C \gamma_\sigma s^b(x) \right) u^c(x) \\ + \left( s^{Ta}(x) C \gamma_\sigma u^b(x) \right) d^c(x) \Big]. \tag{2.4}$$

## B. Correlation functions

Correlation functions at the quark level are obtained through the standard procedure of contracting out pairs of quark fields. Consider the $p \leftrightarrow \Delta^+$ two-point correlation function at the quark level.

$$\langle G_\sigma^{\Delta^+ p}(t; \vec{p}) \rangle = \sum_{\vec{x}} e^{-i\vec{p}\cdot\vec{x}} \langle \Omega \mid T\left(\chi_\sigma^{\Delta^+}(x)\overline{\chi}^p(0)\right) \mid \Omega \rangle, \qquad (2.5)$$

$$= \sqrt{\frac{1}{3}} \sum_{\vec{x}} e^{-i\vec{p}\cdot\vec{x}} \, \epsilon^{abc}\epsilon^{a'b'c'} \Big\{$$
$$2 S_u^{aa'} \, \gamma_5 \, C \, S_d^{Tbb'} \, C \, \gamma_\sigma \, S_u^{cc'}$$
$$- 2 S_d^{aa'} \, \gamma_5 \, C \, S_u^{Tbb'} \, C \, \gamma_\sigma \, S_u^{cc'} \qquad (2.6)$$
$$+ S_u^{aa'} \, tr\left[\gamma_5 \, C \, S_d^{Tbb'} \, C \, \gamma_\sigma \, S_u^{cc'}\right]$$
$$- S_u^{aa'} \, tr\left[\gamma_5 \, C \, S_u^{Tbb'} \, C \, \gamma_\sigma \, S_d^{cc'}\right] \Big\},$$

where the quark-propagator $S_u^{aa'} = S_u^{aa'}(x,0) = T\left(u^a(x), \overline{u}^{a'}(0)\right)$ and similarly for other quark flavors. Here $\Omega$ represents the QCD vacuum.

Similarly, the $\Delta^+ \to p$ correlation function has the following form at the quark level

$$\langle G_\sigma^{p\Delta^+}(t; \vec{p}) \rangle = \sum_{\vec{x}} e^{-i\vec{p}\cdot\vec{x}} \langle \Omega \mid T\left(\chi^p(x)\overline{\chi}_\sigma^{\Delta^+}(0)\right) \mid \Omega \rangle, \qquad (2.7)$$

$$= \sqrt{\frac{1}{3}} \sum_{\vec{x}} e^{-i\vec{p}\cdot\vec{x}} \, \epsilon^{abc}\epsilon^{a'b'c'} \Big\{$$
$$2 S_u^{aa'} \, \gamma_4 \gamma_\sigma \gamma_4 \, C \, S_u^{Tbb'} \, C \, \gamma_5 S_d^{cc'}$$
$$- 2 S_u^{aa'} \, \gamma_4 \gamma_\sigma \gamma_4 \, C \, S_d^{Tbb'} \, C \, \gamma_5 S_u^{cc'} \qquad (2.8)$$
$$+ S_u^{aa'} \, tr\left[\gamma_4 \gamma_\sigma \gamma_4 \, C \, S_u^{Tbb'} \, C \, \gamma_5 \, S_d^{cc'}\right]$$
$$- S_u^{aa'} \, tr\left[\gamma_4 \gamma_\sigma \gamma_4 \, C \, S_d^{Tbb'} \, C \, \gamma_5 \, S_u^{cc'}\right] \Big\},$$

Both of these two-point correlation functions vanish under $SU(2)$-isospin symmetry as required by the isospin invariance of strong interactions. However with $S_d$ replaced by $S_s$ as in $\Sigma^+$ there remains some overlap between the interpolating fields. Unfortunately, this overlap is insufficient to extract any useful information on the spin-1/2 component of the spin-3/2 interpolating field.

The corresponding connected three-point function may be constructed by replacing each of the three propagators $S$, one at a time, by $\hat{S}$ denoting the propagation of a quark in the presence of the electromagnetic current. The three point function analogous to (2.6) is

$$\langle G_\sigma^{\Delta^+ j^\mu p}(t_2, t_1; \vec{p}', \vec{p}) \rangle =$$
$$\sum_{\vec{x}_2, \vec{x}_1} e^{-i\vec{p}'\cdot\vec{x}_2} e^{+i(\vec{p}'-\vec{p})\cdot\vec{x}_1} \langle \Omega \mid T\left(\chi_\sigma^{\Delta^+}(x_2) j^\mu(x_1) \overline{\chi}^p(0)\right) \mid \Omega \rangle, \qquad (2.9)$$
$$= \sqrt{\frac{1}{3}} \sum_{\vec{x}_2} e^{-i\vec{p}'\cdot\vec{x}_2} \, \epsilon^{abc}\epsilon^{a'b'c'} \Big\{$$

$$\begin{aligned}
& 2 S_u^{aa'} \gamma_5 C S_d^{Tbb'} C \gamma_\sigma \widehat{S}_u^{cc'} - 2 S_d^{aa'} \gamma_5 C S_u^{Tbb'} C \gamma_\sigma \widehat{S}_u^{cc'} \\
& + \widehat{S}_u^{aa'} tr\left[\gamma_5 C S_d^{Tbb'} C \gamma_\sigma S_u^{cc'}\right] - \widehat{S}_u^{aa'} tr\left[\gamma_5 C S_u^{Tbb'} C \gamma_\sigma S_d^{cc'}\right] \\
& + 2 \widehat{S}_u^{aa'} \gamma_5 C S_d^{Tbb'} C \gamma_\sigma S_u^{cc'} - 2 S_d^{aa'} \gamma_5 C \widehat{S}_u^{Tbb'} C \gamma_\sigma S_u^{cc'} \\
& + S_u^{aa'} tr\left[\gamma_5 C S_d^{Tbb'} C \gamma_\sigma \widehat{S}_u^{cc'}\right] - S_u^{aa'} tr\left[\gamma_5 C \widehat{S}_u^{Tbb'} C \gamma_\sigma S_d^{cc'}\right] \\
& + 2 S_u^{aa'} \gamma_5 C \widehat{S}_d^{Tbb'} C \gamma_\sigma S_u^{cc'} - 2 \widehat{S}_d^{aa'} \gamma_5 C S_u^{Tbb'} C \gamma_\sigma S_u^{cc'} \\
& + S_u^{aa'} tr\left[\gamma_5 C \widehat{S}_d^{Tbb'} C \gamma_\sigma S_u^{cc'}\right] - S_u^{aa'} tr\left[\gamma_5 C S_u^{Tbb'} C \gamma_\sigma \widehat{S}_d^{cc'}\right] \Big\} \,,
\end{aligned} \quad (2.10)$$

where $\widehat{S}_u^{aa'} = \widehat{S}_u^{aa'}(x_2, 0; t_1, \vec{q}, \mu)$ and $\vec{q} = \vec{p}' - \vec{p}$. The three-point correlation function analogous to (2.8) follows a similar pattern. Using the interpolating fields of (2.3) and (2.4), it is straight forward to verify the $SU(2)$-isospin symmetry relationship for $\Sigma$ three-point transition correlation functions

$$\Sigma^0 = \frac{\Sigma^+ + \Sigma^-}{2}. \quad (2.11)$$

It is useful to examine the symmetries manifest in these correlation functions. First, it is apparent that one of the doubly represented quarks cannot contribute to the transition moments when the masses of the remaining two quarks are equal. In (2.10) the net contribution of the first four terms vanishes under $SU(2)$-isospin symmetry. The contributions of the remaining two quarks to the electromagnetic transition moments differ only by the charges of the quarks and a minus sign. An immediate consequence of this is that all the transition moments for $p \ \gamma \leftrightarrow \Delta^+$ are equal and opposite in sign to the transition moments for $n \ \gamma \leftrightarrow \Delta^0$. Differences between the magnitudes of $n$ and $p$ transition moments reflect isospin-symmetry breaking in the $u$-$d$-quark sector and the contributions of diagrams in which the photon couples to a disconnected quark loop interacting via gluons with the nucleon and $\Delta$. Under $SU(3)$-flavor symmetry, the electromagnetic transition moments for $\Sigma^- \ \gamma \leftrightarrow \Sigma^{*-}$ and for $\Xi^- \ \gamma \leftrightarrow \Xi^{*-}$ vanish. This point was first discussed by Lipkin [21].

Breaking $SU(3)$-flavor symmetry through the introduction of the more massive strange quark allows a nontrivial result for the $\Sigma^-$ and $\Xi^-$ transitions. However, the magnitudes of all the transition moments for $\Sigma^-$ and $\Xi^-$ are governed by the size of the $u$-$s$-quark mass splitting. Furthermore, the transition moments of $\Sigma^- \ \gamma \leftrightarrow \Sigma^{*-}$ and $\Xi^- \ \gamma \leftrightarrow \Xi^{*-}$ will display an approximate symmetry analogous to the exact isospin symmetry of $p \ \gamma \leftrightarrow \Delta^+ = -(n \ \gamma \leftrightarrow \Delta^0)$. The transition moments of $\Sigma^+ \ \gamma \leftrightarrow \Sigma^{*+}$ and $\Xi^0 \ \gamma \leftrightarrow \Xi^{*0}$ will also display a similar approximate symmetry.

The spin-1/2 component of the spin-3/2 hyperon interpolating fields is a source of possible contamination in $\Sigma$ and $\Xi$ transition moments. However, our lattice results for decuplet baryon two-point functions give no evidence of a low-lying spin-1/2 excitation from the spin-1/2 component of the hyperon spin-3/2 interpolating fields. The $\Delta$ baryons are the lowest lying baryons in the mass spectrum having the appropriate quantum numbers and therefore any spin-1/2 excitations have a larger mass and will be exponentially suppressed. The smallness of the two-point correlation functions describing the overlap between octet and decuplet hyperon interpolating fields further supports the absence of any significant octet baryon contaminations in the hyperon correlation functions.

While it may be desirable to use the spin-3/2 projection operator [22]

$$P^{3/2}_{\mu\nu}(p) = g_{\mu\nu} - \frac{1}{3}\gamma_\mu \gamma_\nu - \frac{1}{3p^2}\left(\gamma \cdot p\, \gamma_\mu\, p_\nu + p_\mu\, \gamma_\nu\, \gamma \cdot p\right),  \tag{2.12}$$

one must calculate additional elements of the matrix in Dirac and Lorentz spaces of the three point correlation functions. Such a calculation would exceed our current analysis of 3 Lorentz terms and 2 Dirac terms for two components of the electromagnetic current by a factor of 16.

### III. CORRELATION FUNCTIONS AT THE BARYON LEVEL

#### A. Current Matrix Elements

In this section discussing correlation functions at the hadronic level, the Dirac representation of the $\gamma$-matrices as defined in Bjorken and Drell [23] is used to facilitate calculations of the $\gamma$-matrix algebra. It is then a simple task to account for the differences in $\gamma$-matrix and metric definitions in reporting the final results using Sakurai's notation. In the following we will label octet and decuplet baryons by $N$ and $\Delta$ respectively. However, the results may be applied to any of the octet to decuplet baryon transitions.

The electromagnetic transition moments of $N\,\gamma \leftrightarrow \Delta$ have been thoroughly examined and as a result the current matrix element for even parity transitions is well established. In this investigation we adopt the form which expresses the current matrix element directly in terms of the Sachs transition form factors [24,25]. This matrix element is the most general form required for the description of on-shell nucleon and $\Delta$ states with both real and virtual photon momentum transfers. The current matrix element for $N\,\gamma \to \Delta$ transitions has the following form

$$\langle \Delta(p',s')|j^\mu|N(p,s)\rangle = i\sqrt{\frac{2}{3}}\left(\frac{M_\Delta M_N}{E'_\Delta E_N}\right)^{1/2}\bar{u}_\tau(p',s')\mathcal{O}^{\tau\mu}u(p,s), \tag{3.1}$$

with

$$\mathcal{O}^{\tau\mu} = \mathcal{G}_{M1}(q^2)K^{\tau\mu}_{M1} + \mathcal{G}_{E2}(q^2)K^{\tau\mu}_{E2} + \mathcal{G}_{C2}(q^2)K^{\tau\mu}_{C2}, \tag{3.2}$$

where

$$K^{\tau\mu}_{M1} = -\frac{3}{\{(M_\Delta + M_N)^2 - q^2\}}\epsilon^{\tau\mu\alpha\beta} P_\alpha\, q_\beta\, \frac{(M_\Delta + M_N)}{2M_N}, \tag{3.3a}$$

$$K^{\tau\mu}_{E2} = -K^{\tau\mu}_{M1} - 6\,\Omega^{-1}(q^2)\, \epsilon^{\tau\lambda\alpha\beta}\, P_\alpha\, q_\beta\, \epsilon^{\mu}{}_\lambda{}^{\gamma\delta}\, (2P_\gamma + q_\gamma)\, q_\delta\, i\gamma_5\, \frac{(M_\Delta + M_N)}{2M_N}, \tag{3.3b}$$

$$K^{\tau\mu}_{C2} = -6\,\Omega^{-1}(q^2)\, q^\tau\, \left(q^2 P^\mu - q \cdot P\, q^\mu\right)\, i\gamma_5\, \frac{(M_\Delta + M_N)}{2M_N}, \tag{3.3c}$$

and

$$\Omega(q^2) = \left\{(M_\Delta + M_N)^2 - q^2\right\}\left\{(M_\Delta - M_N)^2 - q^2\right\} \tag{3.4}$$

Momentum is denoted by $p$, $p'$, spin by $s$, $s'$, and $u_\tau(p,s)$ is a spin-vector in the Rarita-Schwinger formalism [26]. Here $q = p' - p$ and $P = (p + p')/2$. The form factors $\mathcal{G}_{M1}$, $\mathcal{G}_{E2}$ and $\mathcal{G}_{C2}$ are referred to as the magnetic dipole $M1$, electric quadrupole $E2$, and electric charge or scalar (time component of the electromagnetic current) quadrupole $C2$ transition form factors [27].

The current matrix element for the inverse reaction $\Delta \to N\gamma$ is defined by the Hermitian conjugate of (3.1),

$$\langle N(p',s')|j^\mu|\Delta(p,s)\rangle \equiv \langle \Delta(p,s)|j^\mu|N(p',s')\rangle^\dagger,$$
$$= -i\sqrt{\frac{2}{3}} \left(\frac{M_\Delta M_N}{E_\Delta E'_N}\right)^{1/2} \overline{u}(p',s')\mathcal{O}^{\tau\mu} u_\tau(p,s), \quad (3.5)$$

where we have used the symmetry

$$\gamma^0 \left(\mathcal{O}^{\tau\mu}\right)^\dagger \gamma^0 = \mathcal{O}^{\tau\mu}. \quad (3.6)$$

### B. Correlation Functions

To isolate and extract the form factors we consider the following ensemble averages of two- and three-point Green functions at the hadronic level

$$\langle G^{NN}(t;\vec{p},\Gamma)\rangle = \sum_{\vec{x}} e^{-i\vec{p}\cdot\vec{x}} \Gamma^{\beta\alpha} \langle \Omega | T\left(\chi^\alpha(x)\overline{\chi}^\beta(0)\right) | \Omega \rangle, \quad (3.7)$$

$$\langle G^{\Delta\Delta}_{\sigma\tau}(t;\vec{p},\Gamma)\rangle = \sum_{\vec{x}} e^{-i\vec{p}\cdot\vec{x}} \Gamma^{\beta\alpha} \langle \Omega | T\left(\chi^\alpha_\sigma(x)\overline{\chi}^\beta_\tau(0)\right) | \Omega \rangle, \quad (3.8)$$

$$\langle G^{\Delta j^\mu N}_\sigma(t_2,t_1;\vec{p}',\vec{p};\Gamma)\rangle =$$
$$\sum_{\vec{x}_2,\vec{x}_1} e^{-i\vec{p}'\cdot\vec{x}_2} e^{+i(\vec{p}'-\vec{p})\cdot\vec{x}_1} \Gamma^{\beta\alpha} \langle \Omega | T\left(\chi^\alpha_\sigma(x_2) j^\mu(x_1) \overline{\chi}^\beta(0)\right) | \Omega \rangle, \quad (3.9)$$

and

$$\langle G^{N j^\mu \Delta}_\sigma(t_2,t_1;\vec{p}',\vec{p};\Gamma)\rangle =$$
$$\sum_{\vec{x}_2,\vec{x}_1} e^{-i\vec{p}'\cdot\vec{x}_2} e^{+i(\vec{p}'-\vec{p})\cdot\vec{x}_1} \Gamma^{\beta\alpha} \langle \Omega | T\left(\chi^\alpha(x_2) j^\mu(x_1) \overline{\chi}^\beta_\sigma(0)\right) | \Omega \rangle. \quad (3.10)$$

Here, $\Gamma$ is a $4 \times 4$ matrix in Dirac space and $\alpha$, $\beta$ are Dirac indices. The subscript $\sigma$ (and $\tau$) is the Lorentz index of the spin-3/2 interpolating field. At the hadronic level one proceeds by inserting a complete set of states $| B(p,s) \rangle$ and defining

$$\langle \Omega | \chi_\sigma(0) | \Delta(p',s') \rangle = Z_\Delta \sqrt{\frac{M_\Delta}{E'_\Delta}} u_\sigma(p',s'), \quad (3.11)$$

where $Z_\Delta$ represents the coupling strength of $\chi(0)$ to the baryon $\Delta$ which may be any baryon resonance having the quantum numbers of the $\Delta$. $E'_\Delta = (\vec{p}'^2 + M_\Delta^2)^{1/2}$ and Dirac indices have been suppressed. Similarly, the overlap between the octet interpolating field and the physical states is defined as

$$\langle \, \Omega \mid \chi(0) \mid N(p,s) \, \rangle = Z_N \sqrt{\frac{M_N}{E_N}} \, u(p,s) \,. \tag{3.12}$$

Using the Rarita-Schwinger spin sum,

$$\sum_s u_\sigma(p,s) \overline{u}_\tau(p,s) = -\frac{\gamma \cdot p + M_\Delta}{2M_\Delta} \left\{ g_{\sigma\tau} - \frac{1}{3}\gamma_\sigma \gamma_\tau - \frac{2 p_\sigma p_\tau}{3 M_\Delta^2} + \frac{p_\sigma \gamma_\tau - p_\tau \gamma_\sigma}{3 M_\Delta} \right\}, \tag{3.13}$$

$$\equiv \Lambda_{\sigma\tau},$$

the Dirac spinor spin sum,

$$\sum_s u(p,s) \overline{u}(p,s) = \frac{\gamma \cdot p + M_N}{2M_N}, \tag{3.14}$$

our usual definitions for $\Gamma$,

$$\Gamma_j = \frac{1}{2} \begin{pmatrix} \sigma_j & 0 \\ 0 & 0 \end{pmatrix} \quad ; \quad \Gamma_0 = \Gamma_4 = \frac{1}{2} \begin{pmatrix} I & 0 \\ 0 & 0 \end{pmatrix}, \tag{3.15}$$

and $\vec{p} = (p, 0, 0)$, the octet and decuplet two-point functions take the following large Euclidean time limits:

$$\langle \, G^{NN}(t; \vec{p}, \Gamma) \, \rangle \simeq Z_N^2 \frac{M_N}{E_N} e^{-E_N t} \, tr \left[ \Gamma \frac{\gamma \cdot p + M_N}{2 M_N} \right], \tag{3.16}$$

$$= Z_N^2 \left( \frac{E_N + M_N}{2 E_N} \right)^{1/2} e^{-E_N t}, \tag{3.17}$$

and

$$\langle \, G^{\Delta\Delta}_{\sigma\tau}(t; \vec{p}, \Gamma) \, \rangle \simeq Z_B^2 \frac{M_\Delta}{E_\Delta} e^{-E_\Delta t} \, tr \left[ \, \Gamma \, \Lambda_{\sigma\tau} \, \right]. \tag{3.18}$$

Similarly, for large Euclidean time separations $t_2 - t_1 \gg 1$ and $t_1 \gg 1$ the three-point function at the hadronic level has the limit

$$\langle \, G^{\Delta j^\mu N}_\sigma(t_2, t_1; \vec{p}', \vec{p}; \Gamma) \, \rangle = \sum_{s,s'} e^{-E'_\Delta (t_2 - t_1)} e^{-E_N t_1}$$

$$\Gamma^{\beta\alpha} \, \langle \, \Omega \mid \chi^\alpha_\sigma \mid \Delta(p',s') \, \rangle \langle \, \Delta(p',s') \mid j^\mu \mid N(p,s) \, \rangle \langle \, N(p,s) \mid \overline{\chi}^\beta \mid \Omega \, \rangle \,. \tag{3.19}$$

where the matrix element of the electromagnetic current is defined in (3.1) through (3.4), and the matrix elements of the interpolating fields are defined by (3.11) and (3.12). Furthermore

$$\langle \, G^{N j^\mu \Delta}_\sigma(t_2, t_1; \vec{p}', \vec{p}; \Gamma) \, \rangle = \sum_{s,s'} e^{-E'_N (t_2 - t_1)} e^{-E_\Delta t_1}$$

$$\Gamma^{\beta\alpha} \, \langle \, \Omega \mid \chi^\alpha \mid N(p',s') \, \rangle \langle \, N(p',s') \mid j^\mu \mid \Delta(p,s) \, \rangle \langle \, \Delta(p,s) \mid \overline{\chi}^\beta_\sigma \mid \Omega \, \rangle, \tag{3.20}$$

where the current matrix element is defined in (3.5).

To isolate and extract the Sachs form factors we construct the following ratio

$$R_\sigma(t_2, t_1; \vec{p}\,', \vec{p}; \Gamma; \mu) =$$
$$\left( \frac{\langle G_\sigma^{\Delta j^\mu N}(t_2, t_1; \vec{p}\,', \vec{p}; \Gamma) \rangle \langle G_\sigma^{N j^\mu \Delta}(t_2, t_1; -\vec{p}, -\vec{p}\,'; \Gamma^\dagger) \rangle}{\langle -g^{\rho\lambda} G_{\rho\lambda}^{\Delta\Delta}(t_2; \vec{p}\,'; \Gamma_4) \rangle \langle G^{NN}(t_2; -\vec{p}; \Gamma_4) \rangle} \right)^{1/2}, \quad (3.21)$$

$$\simeq \left( \frac{E_N + M_N}{2 E_N} \right)^{1/2} \left( \frac{E'_\Delta + M_\Delta}{2 E'_\Delta} \right)^{1/2} \overline{R}_\sigma(\vec{p}\,', \vec{p}; \Gamma; \mu) \quad (3.22)$$

where we have defined the reduced ratio $\overline{R}_\sigma(\vec{p}\,', \vec{p}; \Gamma; \mu)$. Note that there is no implied sum over $\sigma$ in (3.21). For large time separations $\overline{R}_\sigma(\vec{p}\,', \vec{p}; \Gamma; \mu)$ becomes constant and independent of time. In our decuplet baryon analysis we stressed the importance of maintaining the lattice Ward identity when selecting the two-point functions to be used in the analogous ratio. In this case, however, there is no identity to maintain and so we are free to choose any combination of decuplet two-point functions in the ratio. In practise, we use the sum of two-point functions which provides the minimal statistical uncertainties. The optimum combination of decuplet two-point functions used in place of $-g^{\rho\lambda} G_{\rho\lambda}^{\Delta\Delta}(t_2; \vec{p}\,'; \Gamma_4)$ in (3.21) throughout this analysis is

$$-g^{\rho\lambda} G_{\rho\lambda}^{\Delta\Delta}(t_2; \vec{p}\,'; \Gamma_4) = \frac{3}{2} \left\{ G_{22}^{\Delta\Delta}(t_2; \vec{p}\,'; \Gamma_4) + G_{33}^{\Delta\Delta}(t_2; \vec{p}\,'; \Gamma_4) \right\}, \quad (3.23)$$

$$= 2 Z_\Delta^2 \left( \frac{E'_\Delta + M_\Delta}{2 E'_\Delta} \right)^{1/2} e^{-E'_\Delta t_2}. \quad (3.24)$$

In determining the appropriate forms suitable for calculations using Sakurai's conventions the definitions of the $\gamma$-matrices used in the interpolating fields and electromagnetic current are taken into account. The charge form factor is associated with the time component of the electromagnetic current and therefore does not contribute to photoproduction processes. By selecting the time component of the current and $\vec{q} = (q, 0, 0)$ the electric charge transition form factor may be extracted in the following three ways

$$\mathcal{G}_{C2} = 2\sqrt{3} \, \frac{E_N + M_N}{M_\Delta + M_N} \, \frac{M_\Delta M_N}{\vec{q}\,^2} \, \overline{R}_1(\vec{q}, 0; -i\,\Gamma_1; 4), \quad (3.25\text{a})$$

$$\mathcal{G}_{C2} = 4\sqrt{3} \, \frac{E_N + M_N}{M_\Delta + M_N} \, \frac{M_\Delta M_N}{\vec{q}\,^2} \, \overline{R}_2(\vec{q}, 0; i\,\Gamma_2; 4), \quad (3.25\text{b})$$

$$\mathcal{G}_{C2} = 4\sqrt{3} \, \frac{E_N + M_N}{M_\Delta + M_N} \, \frac{M_\Delta M_N}{\vec{q}\,^2} \, \overline{R}_3(\vec{q}, 0; i\,\Gamma_3; 4), \quad (3.25\text{c})$$

where the Green functions at the quark level appearing in $\overline{R}_i$ are defined in terms of Sakurai's notation. While each of (3.25) produces values for $\mathcal{G}_{C2}$ which are in agreement, the statistical uncertainties are relatively large. We will report values for $\mathcal{G}_{C2}$ taken from a fit to the sum of these three ratios.

Similarly, by selecting the spatial component of the current and $\vec{q} = (q, 0, 0)$, the $M1$ and $E2$ transition moments may be isolated. The magnetic $M1$ transition moment may be extracted in the following two ways:

$$\mathcal{G}_{M1} = 2\sqrt{3}\,\frac{E_N + M_N}{M_\Delta + M_N}\,\frac{M_N}{|\vec{q}|}\,\overline{R}_2(\vec{q},0;i\,\Gamma_4;3)\,, \tag{3.26a}$$

$$\mathcal{G}_{M1} = 2\sqrt{3}\,\frac{E_N + M_N}{M_\Delta + M_N}\,\frac{M_N}{|\vec{q}|}\,\left\{\overline{R}_3(\vec{q},0;\Gamma_1;3) - \overline{R}_1(\vec{q},0;\Gamma_3;3)\right\}\,. \tag{3.26b}$$

Values for $\mathcal{G}_{M1}$ taken from either of these ratios are in agreement within statistical uncertainties. Optimum results are obtained from a fit to the sum of these ratios and are reported in the following.

Finally the electric $E2$ transition moment may be obtained from

$$\mathcal{G}_{E2} = \frac{2\sqrt{3}}{3}\,\frac{E_N + M_N}{M_\Delta + M_N}\,\frac{M_N}{|\vec{q}|}\,\left\{\overline{R}_3(\vec{q},0;\Gamma_1;3) + \overline{R}_1(\vec{q},0;\Gamma_3;3)\right\}\,. \tag{3.27}$$

### C. Lattice Techniques

Here we briefly summarize the lattice techniques used in the following calculations. Additional details may be found in Ref. [1]. Wilson's formulation is used for both the gauge and fermionic action. $SU(2)$-isospin symmetry is enforced by equating the Wilson hopping parameters $\kappa_u = \kappa_d = \kappa$. We select three values of $\kappa$, which we denote $\kappa_1 = 0.152$, $\kappa_2 = 0.154$ and $\kappa_3 = 0.156$. To make contact with the physical world, the form factors calculated at our three values of $\kappa$ are linearly extrapolated to $\kappa_{cr}$ where an extrapolation of the squared pion mass vanishes. Differences between linear extrapolations to $m_\pi = 0$ as opposed to the physical pion mass are small and are neglected in the following. To account for the relatively heavy strange quark we fix $\kappa_s = \kappa_1$, the smallest of the three values of $\kappa$ considered. This allows an acceptable extrapolation of the light quarks to the chiral limit through values of quark mass less than or equal to the strange quark mass. Our calculations of octet and decuplet baryon masses indicate that this selection for $\kappa_s$ gives a reasonable description of the strange quark dynamics.

The conserved electromagnetic current is derived from the fermionic action by the Noether procedure. The lattice Ward identity guarantees the lattice electric form factor reproduces the total baryon charge at $q^2 = 0$. The quark propagators coupled with fixed momentum $\vec{q}_1 = (q,0,0)$ to $j^\mu$ are calculated using the sequential source technique (SST) [28–30].

To minimize noise in the Green functions, we exploit the parity symmetry of the correlation functions, and the equal weighting of $\{U\}$ and $\{U^*\}$ gauge configurations in the lattice action. Defining $s_P$ as

$$G(\vec{p}\,',\vec{p},\vec{q};\Gamma) = s_P\,G(-\vec{p}\,',-\vec{p},-\vec{q};\Gamma), \qquad s_P = \pm 1, \tag{3.28}$$

and $s_C$ as

$$\Gamma = s_C\left(\widetilde{C}\Gamma\widetilde{C}^{-1}\right)^*, \qquad s_C = \pm 1, \tag{3.29}$$

where $\widetilde{C} = C\gamma_5$, the correlation functions are real provided

$$s_C = s_P. \tag{3.30}$$

This condition is satisfied with the selections for $\Gamma$ indicated in (3.25a), though (3.27). While this approach requires an extra matrix inversion to determine an additional SST propagator with momentum $-\vec{q}_1$, inclusion of both $\{U\}$ and $\{U^*\}$ configurations in the calculation of the correlation functions provides an unbiased estimate of the ensemble average properties which has substantially reduced fluctuations [31].

Twenty-eight quenched gauge configurations are generated by the Cabibbo-Marinari [32] pseudo-heat-bath method on a $24 \times 12 \times 12 \times 24$ periodic lattice at $\beta = 5.9$. Dirichlet boundary conditions are used for fermions in the time direction. Configurations are selected after 5000 thermalization sweeps from a cold start, and every 1000 sweeps thereafter [33]. Time slices are labeled from 1 to 24, with the $\delta$-function source at $t = 4$. A symmetric combination of the current $(j^\mu(x_1 - \hat{\mu}) + j^\mu(x_1))/2$ is centered at time slice $t_1 = 12$. The spatial direction of the electromagnetic current is chosen in the $z$-direction. The following calculations are done in the lab frame $\vec{p} = 0$, $\vec{p}\,' = \vec{q}_1 = (2\pi/24, 0, 0)$, the minimum nonzero momentum available on our lattice. Using the nucleon mass, the lattice spacing is determined to be $a = 0.128(11)$ fm, $a^{-1} = 1.54(13)$ GeV.

Statistical uncertainties are calculated in a third-order, single elimination jackknife [34,35]. A third order jackknife provides uncertainty estimates for the correlation functions, fits to the correlation functions, and quantities extrapolated to the chiral limit.

Ideally, we would like to calculate the form factors at or very close to $Q^2 = 0$, allowing a direct comparison with the more commonly referred to transition moments. For the decay of our lattice $\Delta$ at rest, energy-momentum conservation requires a photon momentum of approximately 130 MeV, whereas the minimum momentum available on our lattice is 404(35) MeV. Since the minimum momentum available on the lattice is inversely proportional to the longest physical spatial dimension, calculations at $Q^2 = 0$ will require lattices with much larger physical volumes. This difficulty is further compounded by problems associated with tuning the physical lattice size to reproduce the desired momentum or alternatively, using a very fine lattice spacing and much larger lattice volumes to reduce the need for tuning the lattice length.

The momentum transfer at which the form factors are calculated is approximately independent of the baryon under investigation. For $p$, $\Sigma$ and $\Xi$ transitions the $Q^2$ is 0.15(4), 0.16(2) and 0.16(1) GeV$^2$ respectively. A similar independence is seen over the three values of $\kappa$ and $\kappa_{cr}$ under consideration. Therefore variation of the form factor momentum transfer in the extrapolations is not a source of concern.

To make contact with the transition moments at $Q^2 = 0$, we will follow the usual procedure of describing the $Q^2$ dependence of the three transition form factors by a common function [24]. In addition we will assume that the momentum transfer dependence of the transition form factors is similar to the momentum dependence of the decuplet baryon charge form factor $\mathcal{G}_E$. Fortunately, the decuplet baryon elastic form factors are determined at a similar $Q^2$ of 0.16(3) GeV$^2$. This allows a simple scaling of the transition form factors to $Q^2 = 0$ without specific reference to the functional dependence on $Q^2$.

As in our octet and decuplet baryon analyses the scaling is done separately within each quark sector. The $Q^2$ dependence of the individual quark sector contributions can be quite different, particularly in hyperons. Consider, for example, the magnetic transition form factor for hyperons. The strange and light quark sectors are scaled separately by

$$\frac{\mathcal{G}_{M1}^s(0)}{\mathcal{G}_{M1}^s(q^2)} = \frac{\mathcal{G}_E^s(0)}{\mathcal{G}_E^s(q^2)}, \tag{3.31}$$

and similarly for the light quarks, such that the magnetic transition moment of a hyperon is given by

$$\mathcal{G}_{M1}(0) = \mathcal{G}_{M1}^l(0) + \mathcal{G}_{M1}^s(0), \tag{3.32}$$

where $l$ labels the light quarks. For $N\,\gamma \leftrightarrow \Delta$ transitions it is not necessary to separate the $u$- and $d$-quark sectors due to the $SU(2)$-isospin symmetry of the correlation functions. This approach was used in our previous octet and decuplet baryon analyses and is preferred over extrapolations in $q^2$ to $q^2 = 0$ which suffer from large statistical errors.

One might argue that an average of the octet and decuplet baryon charge form factors should be used in scaling the transition moments to $Q^2 = 0$. However differences in the quark charge distribution radii in octet baryons depending on whether the quark is singly or doubly represented would induce an asymmetry in the quark sector contributions to the transition moments at $Q^2 = 0$. This contradicts the symmetry manifest in the three point correlation function of (2.10) which demands the $u$- and $d$-quark contributions to the $p\,\gamma \leftrightarrow \Delta^+$ transition moments to be equal and opposite in sign for equally charged quarks. As a result we choose to use the decuplet baryon charge form factors alone which preserve this symmetry. The difference in the scaled transition moments using the two different approaches is small as the $u$-quark distribution in $p$ is approximately equal to that in $\Delta^+$, while the $d$-quark distribution radius is only slightly larger in $\Delta^+$.

In our decuplet baryon analysis, we found that the charge radius of $\Delta^+$ may actually be smaller than that of the proton [2,17]. This might seem to contradict evidence from the $Q^2$ dependence of transition form factors which suggests the size of the $\Delta$ resonance is larger than the proton [36–38]. We note however, that the predominant difference between the quark charge distributions of the proton and $\Delta$ is the broader distribution of the $d$ quark in $\Delta$. For charge radii the negative charge of the $d$ quark acts to decrease the charge radius of $\Delta$. However, for the transition moments of $p\,\gamma \leftrightarrow \Delta^+$ the $d$ quark contributes in a positive manner. Therefore the distribution radius associated with the transition moment is larger than that associated with electric charge in $\Delta^+$.

## IV. CORRELATION FUNCTIONS

In this section we examine the lattice calculations of the correlation function ratios used to extract the electromagnetic transition form factors. Let us first consider the correlation function ratios used to extract the charge transition form factor, $\mathcal{G}_{C2}$. Figure 1 displays the sum of ratios $\sum_{i=1}^{3} R_i$ for $u$- and $d$-quark contributions to $p\,\gamma \leftrightarrow \Delta^+$ charge transitions. The arguments of $R_i$ are as indicated in equations (3.25). Quark charge factors have not been included. At $\kappa_1$ all quark masses are equal and therefore figure 1 illustrates the quarks contributions to any of the octet to decuplet baryon charge transitions provided doubly represented quarks are identified with the $u$-quark contribution and the singly represented quark with the $d$-quark contribution. The charge transition form factor is small and by time slice 18 the signal is lost in the noise. Correlation functions for lighter quark masses have larger uncertainties.

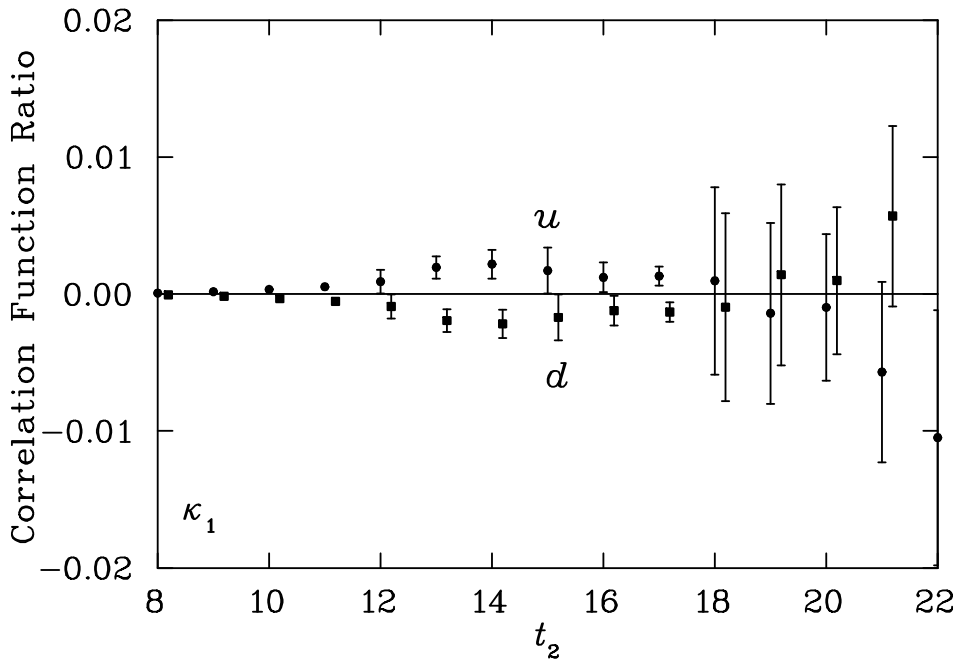

FIG. 1. The ratio sum $\sum_{i=1}^{3} R_i$ for $u$- and $d$-quark contributions to the $p\,\gamma \leftrightarrow \Delta^+$ charge transition form factor. Arguments of $R_i$ are as indicated in equations (3.25). Quark charge factors have not been included. The $d$-quark contributions have been offset in time for clarity.

Quark correlation function ratios for the magnetic transition form factor $\mathcal{G}_{M1}$ of $p\,\gamma \leftrightarrow \Delta^+$ are illustrated in figure 2. The ratios displayed correspond to the sum of the ratios $R_i$ of (3.26a) and (3.26b). Taking results from a sum of all three ratios reduces the statistical uncertainties particularly for larger values of $\kappa$. The time evolution of the correlation function is as follows. At time slice 4, a baryon with the quantum numbers of the proton or $\Delta$ is created. After excited states are exponentially suppressed relative to the baryon ground state, the quarks interact with the electromagnetic current at time slice 12. After a number of time steps, the alternate baryon is annihilated. For large time separations between the electromagnetic current interaction and annihilation, the correlation function ratios are to become constant and independent of time. The correlation function ratios illustrated in figure 2 for our intermediate value of $\kappa$ display small statistical uncertainties. However, the central values do not form as flat a plateau as in our elastic form factor analyses of octet and decuplet baryons. Similar results are seen for each of (3.26a) and (3.26b).

Figure 3 displays the analogous ratios for the magnetic transition form factor of $\Xi$ at the lightest $u$- or $d$-quark masses considered. The strange quark correlation function ratio forms a convincing plateau for time slices greater than or equal to 17. The central values of the light quark correlation function ratio do not form a similar plateau. However for times slices greater than or equal to 17 it is easy to fit the ratio with a horizontal line. This drift in the central values of the light quark correlation function ratios is typical of other baryon transitions such as $N\,\gamma \leftrightarrow \Delta$.

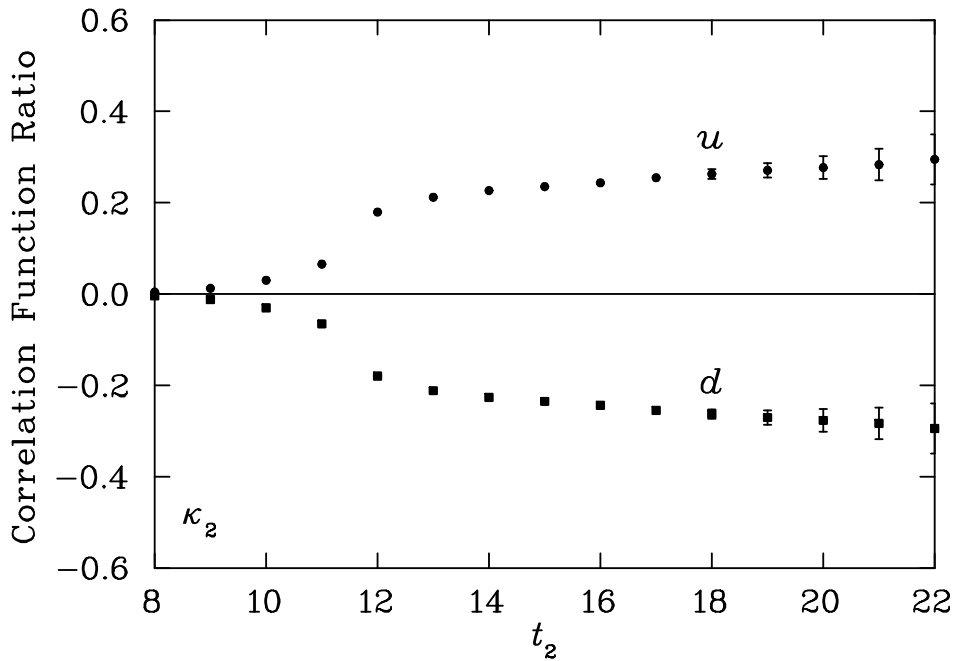

FIG. 2. Quark correlation function ratio sum for the magnetic transition form factor $\mathcal{G}_{M1}$ of $p\,\gamma \leftrightarrow \Delta^+$ at $\kappa_2$. Arguments of $R_i$ are as indicated in equations (3.26).

The form factors are determined by fitting the correlation function ratios by a horizontal line for times $t_2 \gg 12$ and $t_2 \ll 24$. We consider fits of the ratios from time slice 15 through 21 in intervals including 4 to 7 points. The results are selected from these 10 fits based on the flatness of the correlation functions and the statistical uncertainties. Fits of the 4 or 5 points in the time slice interval 17 to 20 or 21 provide the optimum balance between these systematic and statistical uncertainties. This contrasts our study of octet and decuplet baryon elastic form factors where the optimum interval was 16 through 20.

Electric quadrupole transition form factors are determined from the sum of correlation function ratios $R_1$ and $R_3$ as indicated in (3.27). Figure 4 displays $R_1$, $R_3$ and the sum $R_1 + R_3$ for the $u$-quark contribution to the $p\,\gamma \leftrightarrow \Delta$ transition at $\kappa_u = \kappa_1$. In determining $\mathcal{G}_{E2}$ one can fit both $R_1$ and $R_3$ and combine the result or alternatively fit the sum $R_1 + R_3$. The extracted values agree within statistical uncertainties. Since both $R_1$ and $R_3$ should become constant and independent of time we choose to enforce this condition by fitting both $R_1$ and $R_3$ and we refer to these results in the following.

Correlation function ratios for larger values of $\kappa$ have larger statistical uncertainties and therefore the extraction of a clear nonzero value for the electric quadrupole transition moment in this analysis is not possible for most baryons. However a combination of flavor symmetry breaking and the symmetry manifest in the three-point correlation functions allows a prediction for the negatively charged hyperon electric quadrupole moments that differs from zero by two standard deviations.

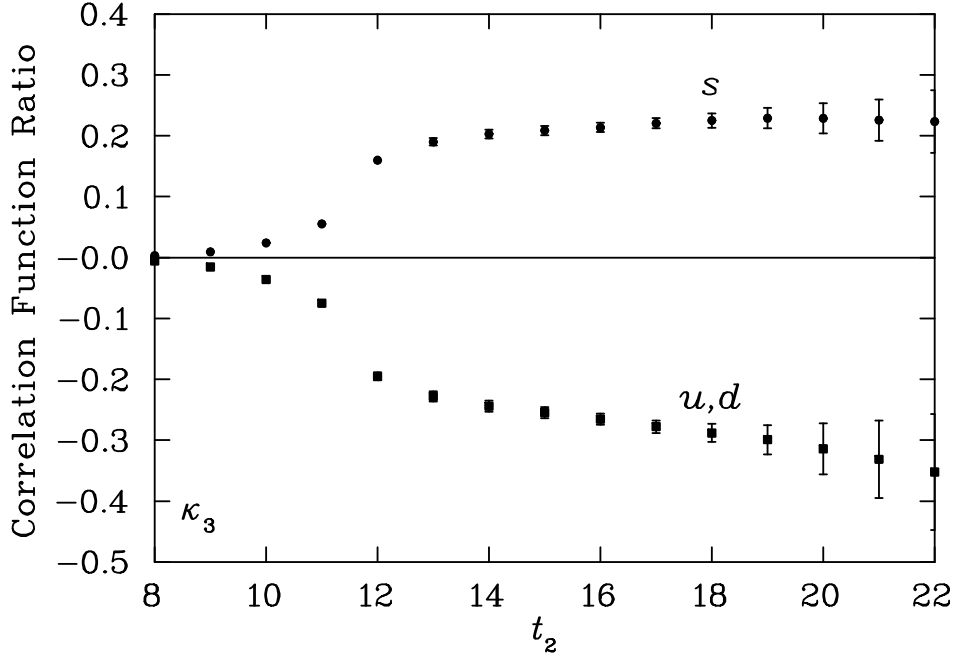

FIG. 3. Quark correlation function ratios for the magnetic transition form factor $\mathcal{G}_{M1}$ of $\Xi\,\gamma \leftrightarrow \Xi^*$ at $\kappa_u = \kappa_d = \kappa_3$. Arguments of $R_i$ are as indicated in equations (3.26).

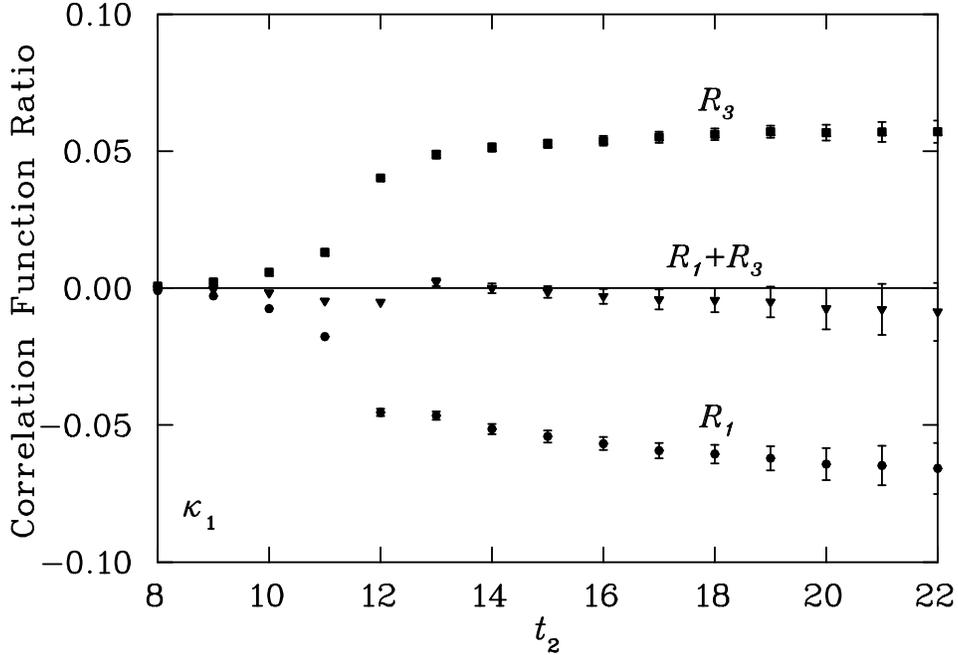

FIG. 4. Correlation function ratios for $u$-quark contributions to the electric quadrupole transition form factor of $p\,\gamma \leftrightarrow \Delta^+$ at $\kappa_u = \kappa_s = \kappa_1$. The arguments of the ratios $R_1$ and $R_3$ are as indicated in (3.27).

The negative charge baryons are unique in that the charge factors multiplying the quark three-point correlation functions are equal ($-1/3$). Moreover the quark correlation functions are equal and opposite in sign in the SU(3) flavor limit. Thus fluctuations in the correlation functions are anticorrelated and to a large extent cancel when added to construct the negative charge baryons.

Figures 5 and 6 display the correlation functions for $s$- and $u$-quark contributions to the $\Xi^0$ $E2$ transition moment at $\kappa_u = \kappa_3$. A comparison with figure 4 for the $p\,\gamma \leftrightarrow \Delta^+$ transition in the SU(3)-flavor regime where $\kappa_u = \kappa_s = \kappa_1$ reveals that SU(3)-flavor symmetry breaking in $\Xi^0\,\gamma \leftrightarrow \Xi^{*0}$ has caused the strange quark contribution to largely vanish while the $u$-quark contribution remains finite. Addition of the quark contributions with anticorrelated fluctuations gives a finite result different from zero by two standard deviations. Similar results are seen for the transition moments of $\Sigma^-\,\gamma \leftrightarrow \Sigma^{*-}$ where the singly represented strange quark contribution is once again seen to largely vanish. The finite $E2$ transition moment has its origin in the light quark which is less localized and more sensitive to the periodic boundary conditions and spatial asymmetries of our lattice.

## V. SIMPLE QUARK MODEL PREDICTIONS

Before proceeding to the lattice results we present here the transition moment predictions of a simple quark model. We have generalized the model calculations of Darewych, Horbatsch and Koniuk [19] to include the entire baryon octet to decuplet transitions. The model is simple in that no attempt has been made to account for configuration mixing [39] in the spin-flavor wave functions or the inclusion of explicit pion dressings of the nucleon [40]. For the baryon transitions under examination here, the general result of their model may be written

$$\mathcal{G}_{M1} = \frac{2\sqrt{2}}{3}\left(\frac{M}{M^*}\right)^{1/2} e^{-K}\left(\mu_D - \mu_S\right) \tag{5.1}$$

where $M$ and $M^*$ are octet and decuplet baryon masses respectively, and $\mu_D$ and $\mu_S$ are the intrinsic moments of the doubly and singly represented quarks respectively. The parameter $K$ is defined as $K = q^2/6\alpha_h^2$ where $q = (M^{*2} - M^2)/2M^*$ and $\alpha_h$ is the harmonic-oscillator strength parameter taken to be 0.41 GeV. For more details of the model calculation, the interested reader is referred to the original publication [19].

It is interesting to note the similarity between (5.1) and (2.10). In the SU(3)-flavor limit, both indicate an equal and opposite weighting of the singly and doubly represented quarks. However with additional information from (2.10) it has become apparent that it is possible to identify one of the doubly represented quarks whose net contribution to the transition moment vanishes in the flavor symmetric limit.

For $\Sigma^0$, the effective moment for $\mu_D$ is $(\mu_u + \mu_d)/2$. In the spirit of the original paper [19] we take $\mu_u = -2\mu_d = 2\mu_p/3$ and $\mu_s/\mu_d = 0.6$. Since configuration mixing in the baryon ground state has not been included, the $E2$ transition moments in this simple model are zero. The magnetic dipole transition moments $\mathcal{G}_{M1}$ are summarized in table I in units of natural magnetons. Helicity amplitudes, decay widths and transition moments using the more widely used conventions of the Particle Data Group [4] are also given. Relationships among these quantities are summarized in the Appendix.

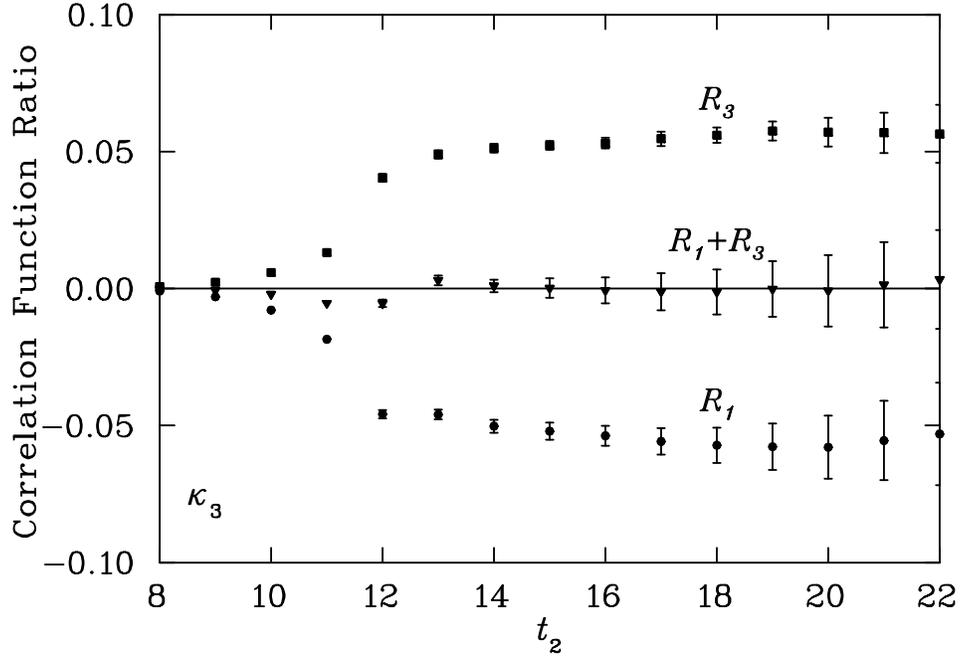

FIG. 5. Correlation function ratios for $s$-quark contributions to the electric quadrupole transition form factor of $\Xi^0\,\gamma \leftrightarrow \Xi^{*0}$ at $\kappa_u = \kappa_3$. The strange quark contribution has largely vanished as the $u$-quark has become light.

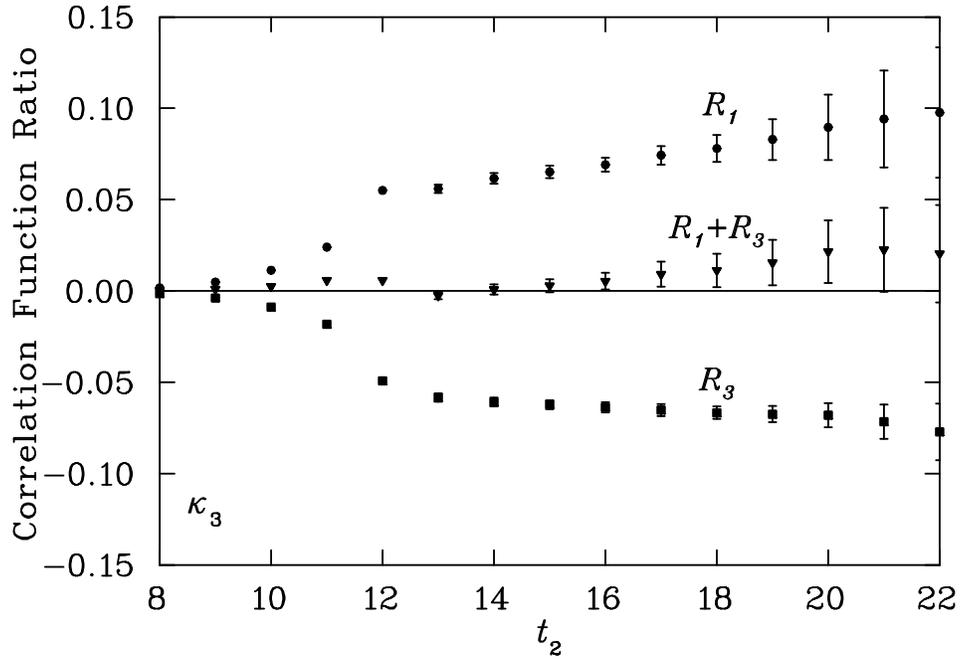

FIG. 6. Correlation function ratios for $u$-quark contributions to the electric quadrupole transition form factor of $\Xi^0\,\gamma \leftrightarrow \Xi^{*0}$ at $\kappa_u = \kappa_3$.

TABLE I. Quark Model Predictions for Transition Moments

| Transition | $\mathcal{G}_{M1}$ ($\mu_B$) | $f_{M1}$ | $A_{1/2}$ | $A_{3/2}$ | $\Gamma$ (keV) |
|---|---|---|---|---|---|
| | | Units of $(10^{-3}$ GeV$^{-1/2})$ | | | |
| $p\,\gamma \leftrightarrow \Delta^+$ | 2.15 | 202 | $-101$ | $-175$ | 330 |
| $n\,\gamma \leftrightarrow \Delta^0$ | $-2.15$ | $-202$ | 101 | 175 | 330 |
| $\Sigma^+\,\gamma \leftrightarrow \Sigma^{*+}$ | 2.61 | 151 | $-75$ | $-131$ | 100 |
| $\Sigma^0\,\gamma \leftrightarrow \Sigma^{*0}$ | 1.10 | 64 | $-32$ | $-55$ | 18 |
| $\Sigma^-\,\gamma \leftrightarrow \Sigma^{*-}$ | $-0.40$ | $-23$ | 12 | 20 | 2.4 |
| $\Xi^0\,\gamma \leftrightarrow \Xi^{*0}$ | $-2.86$ | $-158$ | 79 | 137 | 137 |
| $\Xi^-\,\gamma \leftrightarrow \Xi^{*-}$ | 0.44 | 24 | $-12$ | $-21$ | 3.2 |

For the proton transition moment (5.1) reduces to the well known relation [41]

$$\mu_{p\Delta} = \frac{2\sqrt{2}}{3}\mu_p \tag{5.2}$$

provided one neglects the kinematical factors. For transition moments this is not always a good approximation.

## VI. RESULTS

### A. Magnetic Dipole Transition Form Factors

#### 1. Baryon Transition Moments

Magnetic dipole form factors calculated at $Q^2 \simeq 0.16$ GeV$^2$ are reported in table II at each value of $\kappa$ considered along with the values obtained from a linear extrapolation to $\kappa_{cr}$. Quark sector contributions are reported in table III. Extrapolations of the $M1$ transition moments for a few representative baryons are illustrated in figure 7. The dependence on $\kappa$ is particularly weak for $N\,\gamma \leftrightarrow \Delta$ transitions. SU(3)-flavor symmetry breaking is clearly evident in the $\Sigma^-$ and $\Xi^-$ extrapolations. However, the symmetry of the three point function (2.10) holds to a good approximation even in the broken flavor symmetry regime as the transition moments of $\Sigma^-$ and $\Xi^-$ are roughly equal and opposite in sign. Extrapolated transition moments for the other baryons may be found in table IV.

In the octet baryon analysis it was found that the magnitudes of the lattice results for magnetic moments were consistently smaller than the experimental measurements. It was argued that at $\beta = 5.9$ some deviations from asymptotic scaling may occur. A more recent analysis [15] determines nucleon form factors at $\beta = 6.0$ on a cubic lattice with physical spatial dimensions roughly equal to our smaller $y$ and $z$ dimensions. Some improvement is seen in the magnitudes of the magnetic moments which are still 10 to 15% low compared to experiment. Chiral dressings of the nucleon may cause our linear extrapolation in $1/\kappa$ to underestimate the magnetic transition moments in the physical regime [42]. Finite volume effects may also give rise to the underestimation of the magnetic transition moments as the baryon is restricted by its periodic images. The proton rms electric charge radius at $\kappa_3$ indicates the proton largely fills the lattice in our smaller $y$ and $z$ spatial dimensions. Photon interactions with disconnected quark loops and other non-quenched corrections may also provide additional contributions [43].

TABLE II. Baryon magnetic dipole ($M1$) transition form factors in units of natural magnetons ($\mu_B \equiv e/2M_B$).

| Transition | $\kappa_1 = 0.152$ | $\kappa_2 = 0.154$ | $\kappa_3 = 0.156$ | $\kappa_{cr} = 0.159\,8(2)$ |
|---|---|---|---|---|
| $p\,\gamma \leftrightarrow \Delta^+$ | 1.66(7) | 1.62(8) | 1.61(11) | 1.55(17) |
| $n\,\gamma \leftrightarrow \Delta^0$ | $-1.66(7)$ | $-1.62(8)$ | $-1.61(11)$ | $-1.55(17)$ |
| $\Sigma^+\,\gamma \leftrightarrow \Sigma^{*+}$ | 1.66(7) | 1.67(9) | 1.67(11) | 1.68(15) |
| $\Sigma^0\,\gamma \leftrightarrow \Sigma^{*0}$ | 0.83(4) | 0.80(4) | 0.77(5) | 0.71(6) |
| $\Sigma^-\,\gamma \leftrightarrow \Sigma^{*-}$ | 0.00 | $-0.068(8)$ | $-0.137(23)$ | $-0.258(37)$ |
| $\Xi^0\,\gamma \leftrightarrow \Xi^{*0}$ | $-1.66(7)$ | $-1.69(7)$ | $-1.72(8)$ | $-1.77(11)$ |
| $\Xi^-\,\gamma \leftrightarrow \Xi^{*-}$ | 0.00 | 0.067(7) | 0.132(15) | 0.252(29) |

TABLE III. Quark sector contributions to the $M1$ transition form factor. Quantities are normalized to unit charge and are reported in units of natural magnetons ($\mu_B \equiv e/2M_B$).

| Transition | Quark | $\kappa_1 = 0.152$ | $\kappa_2 = 0.154$ | $\kappa_3 = 0.156$ | $\kappa_{cr} = 0.159\,8(2)$ |
|---|---|---|---|---|---|
| $p\,\gamma \leftrightarrow \Delta^+$ | $u$ | 1.66(7) | 1.62(8) | 1.61(11) | 1.55(17) |
|  | $d$ | $-1.66(7)$ | $-1.62(8)$ | $-1.61(11)$ | $-1.55(17)$ |
| $\Sigma^+\,\gamma \leftrightarrow \Sigma^{*+}$ | $u$ | 1.66(7) | 1.73(9) | 1.81(12) | 1.94(19) |
|  | $s$ | $-1.66(7)$ | $-1.53(8)$ | $-1.40(9)$ | $-1.16(10)$ |
| $\Xi^0\,\gamma \leftrightarrow \Xi^{*0}$ | $u$ | $-1.66(7)$ | $-1.76(8)$ | $-1.85(8)$ | $-2.02(12)$ |
|  | $s$ | 1.66(7) | 1.56(7) | 1.45(8) | 1.26(12) |

TABLE IV. Baryon $M1$ transition form factors at $Q^2 = 0$.

| Transition | $\kappa_1 = 0.152$ | $\kappa_2 = 0.154$ | $\kappa_3 = 0.156$ | $\kappa_{cr} = 0.159\,8(2)$ |
|---|---|---|---|---|
| $p\,\gamma \leftrightarrow \Delta^+$ | 1.94(8) | 1.95(8) | 1.97(12) | 1.99(20) |
| $n\,\gamma \leftrightarrow \Delta^0$ | $-1.94(8)$ | $-1.95(8)$ | $-1.97(12)$ | $-1.99(20)$ |
| $\Sigma^+\,\gamma \leftrightarrow \Sigma^{*+}$ | 1.94(8) | 1.98(10) | 2.03(14) | 2.11(21) |
| $\Sigma^0\,\gamma \leftrightarrow \Sigma^{*0}$ | 0.97(4) | 0.94(4) | 0.92(5) | 0.86(8) |
| $\Sigma^-\,\gamma \leftrightarrow \Sigma^{*-}$ | 0.00 | $-0.099(13)$ | $-0.20(4)$ | $-0.38(6)$ |
| $\Xi^0\,\gamma \leftrightarrow \Xi^{*0}$ | $-1.94(8)$ | $-2.02(8)$ | $-2.10(8)$ | $-2.24(11)$ |
| $\Xi^-\,\gamma \leftrightarrow \Xi^{*-}$ | 0.00 | 0.098(11) | 0.202(28) | 0.37(5) |

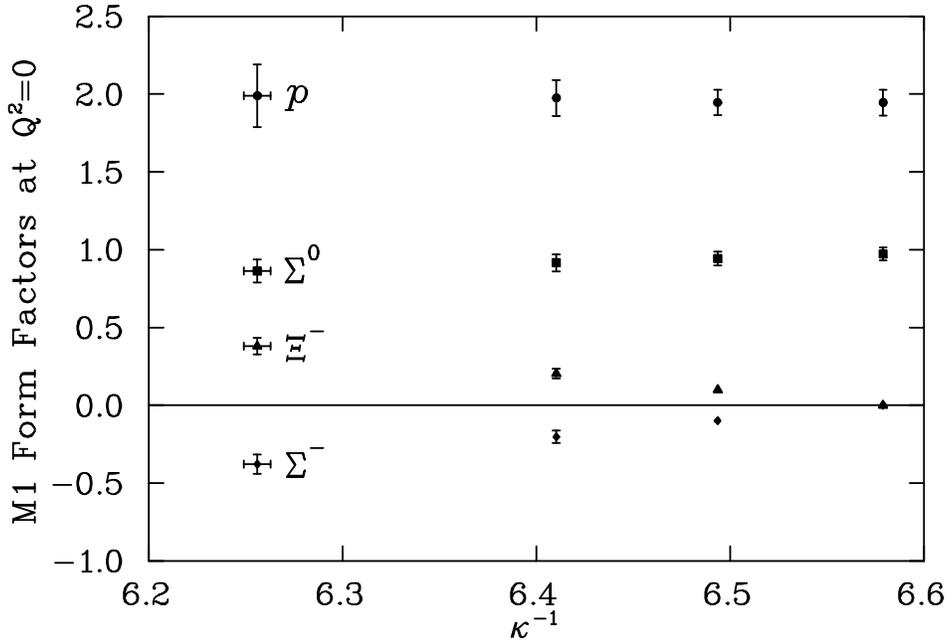

FIG. 7. Extrapolation of $M1$ transition moments for a few baryons representative of baryon octet to decuplet transitions. The dependence on $\kappa$ is particularly weak for $N\,\gamma \leftrightarrow \Delta$ transitions.

To reduce the effects of these uncertainties, ratios of the lattice results to the lattice proton result are used when making comparisons with experimental measurements or model calculations. Table V reports the ratio of the extrapolated baryon magnetic dipole transition moments to the proton magnetic moment scaled to reproduce the proton magnetic moment. Values are given for the Sachs form factor, $\mathcal{G}_{M1} = \mu/\mu_B$, where $\mu_B$ is the unit of natural magnetons ($e/2M_B$, $M_B$ is the mass of the octet baryon). Values are also reported for the Particle Data Group [4] convention, $f_{M1}$, calculated from $\mathcal{G}_{M1}$ using the physical baryon masses [44]. Relationships for the Sachs form factors and the conventions of the Particle Data Group are summarized in the appendix. The SU(3)-flavor symmetry relationships $(\Sigma^-\,\gamma \leftrightarrow \Sigma^{*-}) = -(\Xi^-\,\gamma \leftrightarrow \Xi^{*-})$ and $(\Sigma^+\,\gamma \leftrightarrow \Sigma^{*+}) = -(\Xi^0\,\gamma \leftrightarrow \Xi^{*0})$ are seen to hold to a good approximation. This suggests the quark contributions to the transition moments do not depend strongly on the baryon in which the quarks reside.

In figure 8 the lattice predictions of the Sachs form factor $\mathcal{G}_{M1}$ are compared with those of the simple quark model reviewed in the previous section. Remarkable agreement is seen throughout the baryon octet to decuplet transitions. A similar agreement was seen between the lattice results and the simple quark model in our decuplet baryon analysis. Further examination revealed that the agreement was largely due to an approximate baryon inde-pendence of the quark effective magnetic moments. This was in contrast to our octet baryon analysis where it was found that the electromagnetic properties of a quark have a strong dependence on the baryon in which it resides.

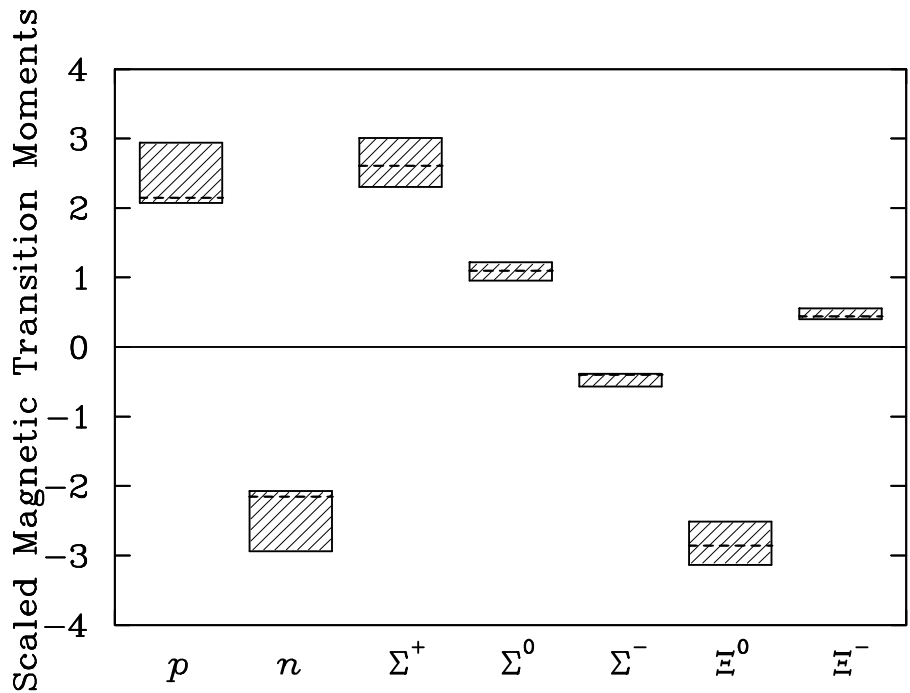

FIG. 8. Lattice predictions for the Sachs form factor $\mathcal{G}_{M1}$. The dashed lines are predictions based on the simple quark model reviewed in section V. Remarkable agreement is seen throughout the baryon octet to decuplet transitions.

TABLE V. Baryon $M1$ transition moments. The lattice results have been scaled to reproduce the proton magnetic moment.

| Transition | $\mathcal{G}_{M1}$ | $f_{M1}$ |
| | ($\mu_B$) | (GeV$^{-1/2}$) |
| --- | --- | --- |
| $p\gamma \leftrightarrow \Delta^+$ | 2.46(43) | 0.231(41) |
| $n\gamma \leftrightarrow \Delta^0$ | −2.46(43) | −0.231(41) |
| $\Sigma^+\gamma \leftrightarrow \Sigma^{*+}$ | 2.61(35) | 0.151(20) |
| $\Sigma^0\gamma \leftrightarrow \Sigma^{*0}$ | 1.07(13) | 0.062(8) |
| $\Sigma^-\gamma \leftrightarrow \Sigma^{*-}$ | −0.47(9) | −0.027(5) |
| $\Xi^0\gamma \leftrightarrow \Xi^{*0}$ | −2.77(31) | −0.153(17) |
| $\Xi^-\gamma \leftrightarrow \Xi^{*-}$ | 0.47(8) | 0.026(4) |

## 2. Effective Quark Moments

It is interesting to examine the electromagnetic contributions of the quark sectors individually. The baryon transition moments are composed of a sum of quark sector contributions and interesting phenomena may remain hidden in taking the sum. To define an effective quark moment we turn to the quark model reviewed in the previous section. Defining $\mathcal{G}_{M1}^{D}$ ($\mathcal{G}_{M1}^{S}$) to be the doubly (singly) represented quark sector contribution to the transition moment, we define the effective quark moments to be

$$\mu_D^B = +\frac{3\sqrt{2}}{4} \left(\frac{M_B^*}{M_B}\right)^{1/2} \left(\frac{M_N}{M_B}\right) \mathcal{G}_{M1}^{D}, \qquad (6.1a)$$

$$\mu_S^B = -\frac{3\sqrt{2}}{4} \left(\frac{M_B^*}{M_B}\right)^{1/2} \left(\frac{M_N}{M_B}\right) \mathcal{G}_{M1}^{S}, \qquad (6.1b)$$

where $M_B$ and $M_B^*$ are the masses of the octet and decuplet baryons under transition. The second mass ratio $M_N/M_B$ expresses the form factor $\mathcal{G}_{M1}$ in units of nuclear magnetons. For the central values of the lattice masses, the factor $\exp(-K)$ appearing in (5.1) takes the values $0.986 \pm 0.003$. Since this factor is approximately one and has an obvious model dependence we have dropped this factor from the definition of the effective quark moment.

The three-point correlation function of (2.10) indicates the quark sector contributions to the transition moment are equal and opposite under SU(3) symmetry. A similar symmetry was seen in the decuplet baryon three-point correlation functions. Hence an interesting question to ask is whether SU(3)-flavor symmetry is broken in the same manner in the magnetic transition moments as in the magnetic moments of the decuplet baryons. This question may be answered with minimal model dependence by taking a ratio of the effective quark transition moments and comparing the result with a similar ratio of effective quark magnetic moments in decuplet baryons. In this way the mass ratios appearing in (6.1) and other factors neglected in the effective quark transition moment definitions are eliminated from the SU(3) flavor symmetry breaking measure. To compare SU(3)-flavor symmetry breaking we calculate the following ratio

$$(\mu_s/\mu_u)_{\text{transitions}} \Big/ (\mu_s/\mu_u)_{\text{decuplet}} . \qquad (6.2)$$

Of course a simple quark model defines this ratio to be 1. The lattice results indicate that SU(3)-flavor symmetry is broken in a different manner for the transition moments compared to the decuplet elastic moments. For the transition $\Sigma\,\gamma \leftrightarrow \Sigma^*$ the ratio of (6.2) is 0.82(10). Similar results hold for $\Xi\,\gamma \leftrightarrow \Xi^*$ transitions where the ratio is 0.83(9).

To discover whether it is a suppression of the strange-quark or an enhancement of the $u$-quark contributions to the transition moments that is responsible for the deviations from the simple quark model description of flavor symmetry breaking, we turn to the actual values of the effective quark moments. Figure 9 illustrates the effective quark moments defined in (6.1) normalized to unit charge. To a good approximation, the effective quark moments are independent of the environment in which the quark resides, with the possible exception of the $u$ quark in $\Xi$. The ratio of effective $u$-quark transition moments for $\Xi$ and $p$ indicate a 15(7)% suppression of the effective $u$-quark moment in $\Xi^0 \; \gamma \leftrightarrow \Xi^{*0}$ transitions. A similar ratio for the $s$-quarks in $\Sigma$ and $\Xi$ transitions indicates $s_\Xi$ may be smaller by 10(10)%.

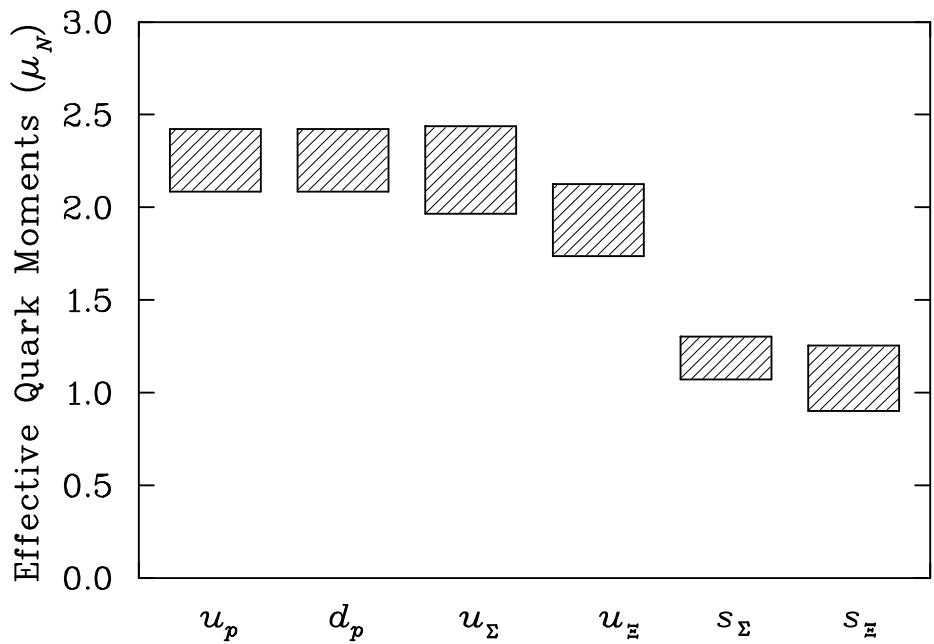

FIG. 9. Effective quark moments determined from the quark sector contributions to radiative transitions of octet and decuplet baryons. The quark moments are defined in (6.1), and are normalized to unit charge. To a good approximation, the effective quark moments are independent of the environment in which the quark resides, with the possible exception of the $u$ quark in $\Xi$.

In our decuplet baryon analysis a similar gentle environment sensitivity of the quark moments was seen. Figure 10 summarizes the effective quark moments for quarks in decuplet baryons. Here, the quark moments decrease as strangeness is added. This may be attributed to the role the baryon mass plays in setting the scale at which the quarks contribute to the magnetic moment.

Similar conclusions could be drawn for the transition moments if it were not for the lack of suppression in the $u$-quark moment in $\Sigma^+$ transitions. However the $u$-quark in $\Sigma^+$ is doubly represented and SU(3)-flavor symmetry is broken. Terms of the three-point transition correlation function of (2.10) which cannot contribute to nucleon transitions can now provide additional contributions to the $u$-quark sector of $\Sigma^+$ transitions. Similar effects may be occurring in the effective quark transition moment $s_\Xi$. However, the influence of an additional strange quark in the baryon appears to be playing a stronger role, and decreases the magnitude of the magnetic moment contribution.

Figure 11 displays ratios of the effective quark moments from transitions to the effective quark moments from decuplet baryons. The light quark transition moments are more consistent with the values defined by decuplet baryon magnetic moments than the strange quark transition moments. Hence, it is a suppression of the effective transition moments of strange quarks relative to their values in decuplet baryons that is largely responsible for the deviations from the simple quark model predictions of flavor symmetry breaking. The effective strange-quark transition moments are suppressed by 20(9)% and 26(10)% relative to their decuplet baryon values for $\Sigma$ and $\Xi$ transitions respectively.

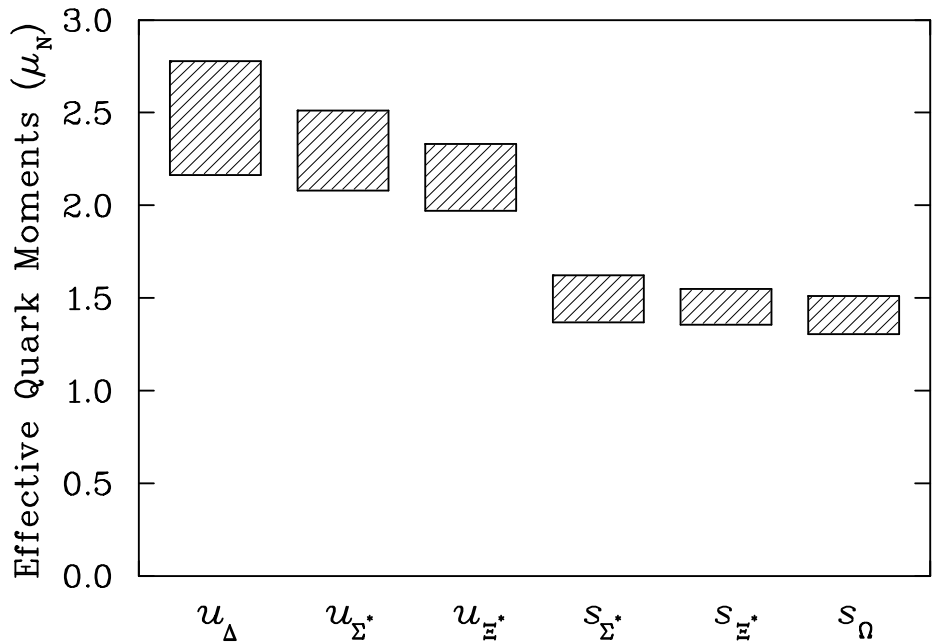

FIG. 10. Effective quark moments determined from the quark sector contributions to decuplet baryon magnetic moments. Approximate baryon independence of the quark moments is displayed. However, some decrease is seen in the effective quark moments as strangeness is added.

In our previous analyses, we have found the effective moment of the $u$ quark in the proton to be equal, to a good approximation, to the $u$-quark moment in the $\Delta$. Similarly the effective $u$-quark moment determined from transitions of $N\ \gamma \leftrightarrow \Delta$ is in agreement with the effective $u$-quark moments in $p$ and $\Delta$. Given the similarity of $u$ quark properties in $p$ and $\Delta$, this is as one might expect. Therefore the drift in the central values of the three-point correlation functions may simply be an indication of the need for better statistics as opposed to a signature of the correlation functions failing to reach the plateau region before the lattice boundary is encountered.

Finally it is worth commenting on why the lattice results suggest values for the $M1$ transition moments of $N\ \gamma \leftrightarrow \Delta$ that are somewhat larger than that anticipated by the simple quark model. The main source of the difference stems from the manner in which the quarks contribute to the magnetic moment of the proton, which has been used to set the overall scale of the magnetic moments. While SU(6)-spin-flavor symmetry predicts

$$\mu_p = \frac{4}{3}\mu_u - \frac{1}{3}\mu_d\,, \qquad (6.3)$$

our octet baryon analysis indicates that the $d$-quark contribution is suppressed in the lattice results by a factor of approximately 2 from that anticipated by SU(6). Since the effective $u$-quark moment in the proton is approximately equal to the effective $u$-quark moment in $p\,\gamma \leftrightarrow \Delta^+$ transitions, the difference in lattice and quark model predictions for the $\mu_{p\Delta}/\mu_p$ ratio may be attributed to the smallness of the $d$-quark moment contribution to the proton magnetic moment.

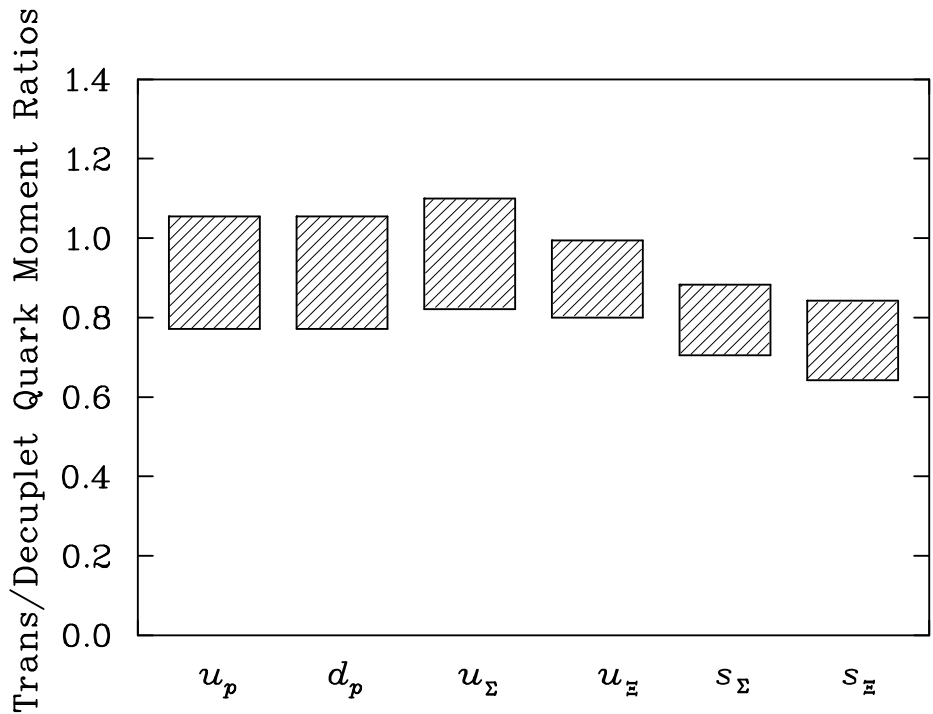

FIG. 11. Ratios of the effective quark moments from transitions to the effective quark moments from decuplet baryons. The light quark transition moments are more consistent with the values defined by decuplet baryon magnetic moments than the strange quark transition moments.

### B. Electric Charge Transition Form Factors

Electric charge form factors calculated at $Q^2 \simeq 0.16$ GeV$^2$ are reported in table VI. The statistical uncertainties are large and prevent us from drawing any strong conclusions. However, it interesting to note that the central value of the charge form factor for the $p\,\gamma \leftrightarrow \Delta^+$ transition has the sign and magnitude anticipated by multipole analyses [36,37] at about $-10\%$ of the magnetic form factor.

The lattice results suggest the charge form factor may be large for the negatively charged hyperon transitions. This is due to an addition of the quark sector contributions which are illustrated in table VII. Examination of the correlation functions for the hyperons reveals that the correlation functions are somewhat noisy and do not form a convincing plateau. A future high statistics analysis may provide some interesting insights into the charge form factor.

TABLE VI. Scalar Quadrupole ($C2$) transition form factors.

| Transition | $\kappa_1 = 0.152$ | $\kappa_2 = 0.154$ | $\kappa_3 = 0.156$ | $\kappa_{cr} = 0.1598(2)$ |
|---|---|---|---|---|
| $p\,\gamma \leftrightarrow \Delta^+$ | 0.04(7) | −0.02(16) | −0.04(26) | −0.13(41) |
| $n\,\gamma \leftrightarrow \Delta^0$ | −0.04(7) | 0.02(16) | 0.04(26) | 0.13(41) |
| $\Sigma^+\,\gamma \leftrightarrow \Sigma^{*+}$ | 0.04(7) | 0.02(9) | 0.15(31) | 0.06(38) |
| $\Sigma^0\,\gamma \leftrightarrow \Sigma^{*0}$ | 0.02(3) | −0.04(7) | −0.05(15) | −0.17(14) |
| $\Sigma^-\,\gamma \leftrightarrow \Sigma^{*-}$ | 0.00 | −0.11(12) | −0.25(30) | −0.44(54) |
| $\Xi^0\,\gamma \leftrightarrow \Xi^{*0}$ | −0.04(7) | 0.01(12) | 0.00(15) | 0.06(19) |
| $\Xi^-\,\gamma \leftrightarrow \Xi^{*-}$ | 0.00 | 0.09(12) | 0.25(11) | 0.46(33) |

TABLE VII. Quark sector contributions to the $C2$ transition form factor. Quantities are normalized to unit charge.

| Transition | Quark | $\kappa_1 = 0.152$ | $\kappa_2 = 0.154$ | $\kappa_3 = 0.156$ | $\kappa_{cr} = 0.1598(2)$ |
|---|---|---|---|---|---|
| $p\,\gamma \leftrightarrow \Delta^+$ | $u$ | 0.04(7) | −0.02(16) | −0.04(26) | −0.13(41) |
|  | $d$ | −0.04(7) | 0.02(16) | 0.04(26) | 0.13(41) |
| $\Sigma^+\,\gamma \leftrightarrow \Sigma^{*+}$ | $u$ | 0.04(7) | 0.13(16) | 0.4(5) | 0.5(8) |
|  | $s$ | −0.04(7) | 0.19(24) | 0.4(5) | 0.8(7) |
| $\Xi^0\,\gamma \leftrightarrow \Xi^{*0}$ | $u$ | −0.04(7) | −0.08(8) | −0.26(17) | −0.34(27) |
|  | $s$ | 0.04(7) | −0.20(35) | −0.51(29) | −1.0(8) |

### C. Electric Quadrupole Transition Form Factors

As discussed in section IV, only two of the octet to decuplet $E2$ transition form factors are statistically different from zero. Figure 12 illustrates the extrapolation of these $E2$ transition form factors for $\Sigma^-$ and $\Xi^-$ transitions. At the largest value of quark mass considered, SU(3)-flavor symmetry is exact and the transition moments vanish for these baryons. The symmetry of the three-point correlation function continues to hold even in the broken flavor-symmetry region as the $E2$ transition form factors are seen to be approximately equal and opposite. Values for the $E2$ transition form factors at each value of $\kappa$ considered as well as at $\kappa_{cr}$ are summarized in table VIII. Quark contributions to these form factors are given in table IX.

TABLE VIII. Baryon electric quadrupole ($E2$) transition form factors.

| Transition | $\kappa_1 = 0.152$ | $\kappa_2 = 0.154$ | $\kappa_3 = 0.156$ | $\kappa_{cr} = 0.1598(2)$ |
|---|---|---|---|---|
| $p\,\gamma \leftrightarrow \Delta^+$ | −0.021(21) | −0.032(45) | −0.03(9) | −0.05(13) |
| $n\,\gamma \leftrightarrow \Delta^0$ | 0.021(21) | 0.032(45) | 0.03(9) | 0.05(13) |
| $\Sigma^+\,\gamma \leftrightarrow \Sigma^{*+}$ | −0.021(21) | −0.034(37) | −0.06(6) | −0.08(10) |
| $\Sigma^0\,\gamma \leftrightarrow \Sigma^{*0}$ | −0.011(11) | −0.014(18) | −0.02(3) | −0.03(5) |
| $\Sigma^-\,\gamma \leftrightarrow \Sigma^{*-}$ | 0.000 | 0.006(3) | 0.016(11) | 0.024(12) |
| $\Xi^0\,\gamma \leftrightarrow \Xi^{*0}$ | 0.021(21) | 0.026(27) | 0.032(35) | 0.04(5) |
| $\Xi^-\,\gamma \leftrightarrow \Xi^{*-}$ | 0.000 | −0.005(2) | −0.014(6) | −0.022(9) |

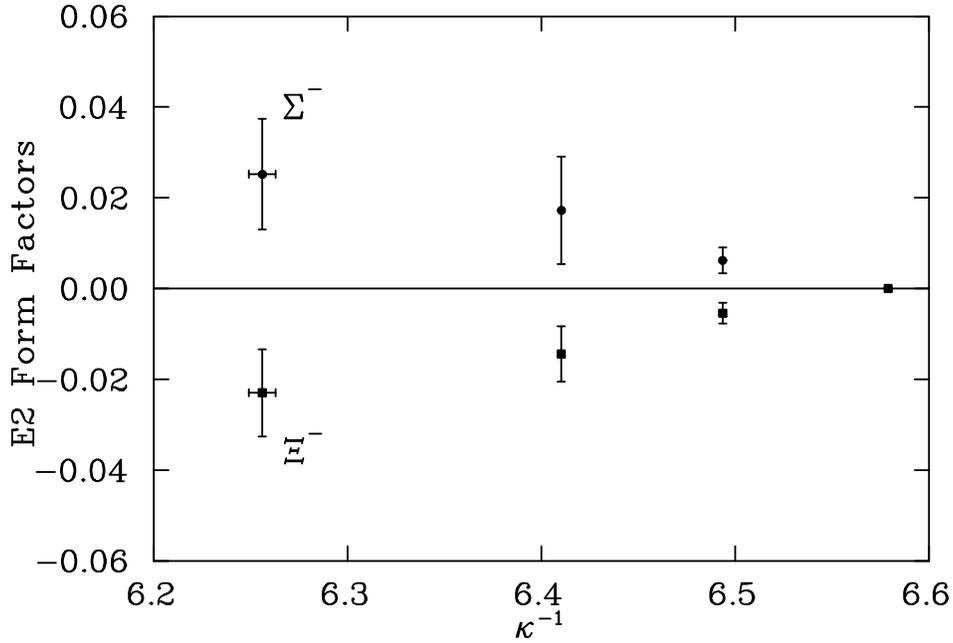

FIG. 12. Extrapolation of the $E2$ transition form factors for $\Sigma^-$ and $\Xi^-$ transitions. The symmetry of the three-point correlation function continues to hold even in the broken flavor-symmetry region as the $E2$ transition form factors are seen to be approximately equal and opposite.

The quantity that has captured the attention and excitement of the field is the $E2/M1$ ratio of electromagnetic form factors $\mathcal{R}_{EM} \equiv f_{E2}/f_{M1} = -\mathcal{G}_{E2}/\mathcal{G}_{M1}$. Analyses of experimental data place this ratio at $-(1.57 \pm 0.72)\%$, [45]; $-(1.5 \pm 0.4)\%$, [46]; $-3.1\%$, [47]; and $+(3.7 \pm 0.4)\%$, [3]. The reader is directed to the original papers for a complete discussion of the uncertainties reflected in the numerical error bars given here. Details of the lattice determinations of $\mathcal{R}_{EM}$ are reported in table X. A summary of the lattice ratios $\mathcal{R}_{EM}$ is given in figure 13. Unfortunately the statistical uncertainties in the lattice results are relatively large. All the results from the experimental analyses lie comfortably within our determination from first principles of $+(3 \pm 8)\%$.

The interesting feature, however, is that the lattice approach can make predictions for the $E2$ transition moments of $\Sigma^-$ and $\Xi^-$ transitions which will provide vital information to those developing models. Still, the present values must be taken with some caution. The non-vanishing contribution to these $E2$ moments has its origin in the light $d$-quark sector. This quark has a broad distribution radius which may be sensitive to the asymmetries our elongated lattice. Interactions with periodic images may induce an $E2$ moment. It is clear that a future calculation on a cubic lattice will provide much needed insight into the $E2$ transition form factor.

TABLE IX. Quark sector contributions to the $E2$ transition form factor. Quantities are normalized to unit charge.

| Transition | Quark | $\kappa_1 = 0.152$ | $\kappa_2 = 0.154$ | $\kappa_3 = 0.156$ | $\kappa_{cr} = 0.1598(2)$ |
|---|---|---|---|---|---|
| $p\,\gamma \leftrightarrow \Delta^+$ | $u$ | $-0.021(21)$ | $-0.032(45)$ | $-0.03(9)$ | $-0.05(13)$ |
| | $d$ | $0.021(21)$ | $0.032(45)$ | $0.03(9)$ | $0.05(13)$ |
| $\Sigma^+\,\gamma \leftrightarrow \Sigma^{*+}$ | $u$ | $-0.021(21)$ | $-0.040(38)$ | $-0.07(7)$ | $-0.11(10)$ |
| | $s$ | $0.021(21)$ | $0.022(36)$ | $0.02(7)$ | $0.03(10)$ |
| $\Xi^0\,\gamma \leftrightarrow \Xi^{*0}$ | $u$ | $0.021(21)$ | $0.032(28)$ | $0.05(4)$ | $0.07(5)$ |
| | $s$ | $-0.021(21)$ | $-0.016(26)$ | $-0.00(4)$ | $0.01(5)$ |

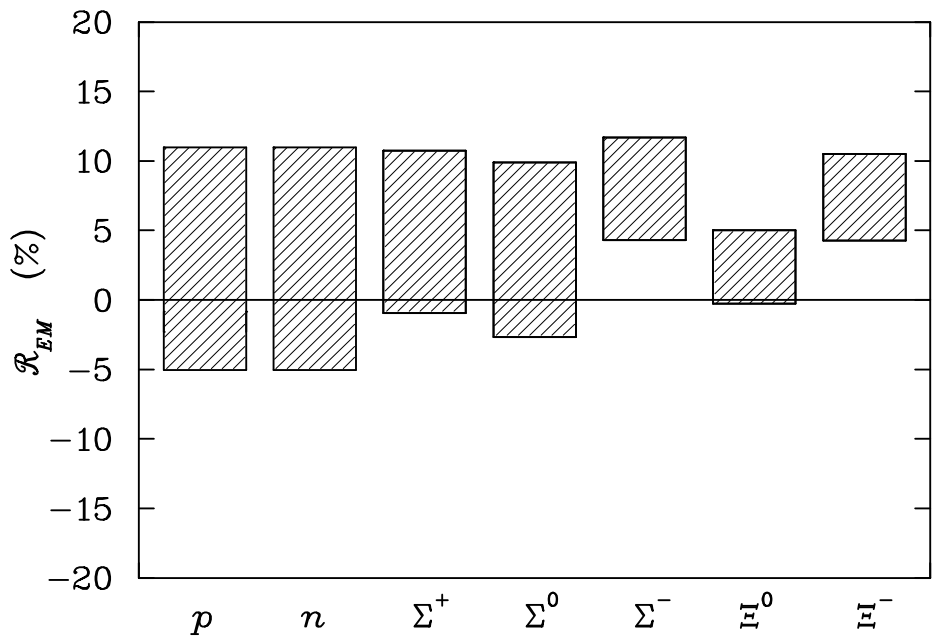

FIG. 13. A summary of lattice calculations of the $E2/M1$ ratio $\mathcal{R}_{EM}$. Statistical uncertainties in the lattice results are too large to favor any particular model calculation.

TABLE X. The $E2/M1$ ratio $\mathcal{R}_{EM}$ at $Q^2 = 0$.

| Transition | $\kappa_1 = 0.152$ | $\kappa_2 = 0.154$ | $\kappa_3 = 0.156$ | $\kappa_{cr} = 0.159\,8(2)$ |
|---|---|---|---|---|
| $p\,\gamma \leftrightarrow \Delta^+$ | 0.013(13) | 0.019(27) | 0.02(5) | 0.03(8) |
| $n\,\gamma \leftrightarrow \Delta^0$ | 0.013(13) | 0.019(27) | 0.02(5) | 0.03(8) |
| $\Sigma^+\,\gamma \leftrightarrow \Sigma^{*+}$ | 0.013(13) | 0.021(22) | 0.03(4) | 0.05(6) |
| $\Sigma^0\,\gamma \leftrightarrow \Sigma^{*0}$ | 0.013(13) | 0.018(23) | 0.03(4) | 0.04(6) |
| $\Sigma^-\,\gamma \leftrightarrow \Sigma^{*-}$ | | 0.076(34) | 0.10(7) | 0.08(4) |
| $\Xi^0\,\gamma \leftrightarrow \Xi^{*0}$ | 0.013(13) | 0.016(16) | 0.019(20) | 0.024(27) |
| $\Xi^-\,\gamma \leftrightarrow \Xi^{*-}$ | | 0.067(27) | 0.087(37) | 0.074(30) |

## VII. MODEL COMPARISONS

Lattice predictions for ratios of octet baryon magnetic moments to the lattice proton moment are in excellent agreement with experimental measurements when the baryon moment is positive. Ratios for baryons with negative magnetic moments are more sensitive to contributions from disconnected quark loop contributions which are not included in present form factor calculations [43]. Since the transition moment for $p\,\gamma \leftrightarrow \Delta^+$ is positive, the lattice approach should provide reliable estimates of the magnetic transition moment ratio $\mu_{p\Delta^+}/\mu_p$.

Figure 14 summarizes many calculations of the magnetic transition moment, $f_{M1}$. The need for a calculation of this quantity from first principles is reflected in the wide range of values for this quantity. The calculations have been categorized into six different approaches including analyses of experimental pion photoproduction data (Expt.), our lattice QCD calculation (Latt.), nonrelativistic quark model determinations (Q.M.), hedgehog models including the Skyrme and Hybrid models (Hedge.), bag models (Bag), and a Bethe-Salpeter calculation (B.S.). In the following we discuss each of these approaches in relation to our new lattice determination, and give specific references to the model determinations. These results are by no means exhaustive, but are representative and indicate the breadth of interest in the determination of the transition moment.

The analyses of experimental data include (from top down) Davidson, Mukhopadhyay, and Wittman's investigation using an effective Lagrangian with a number of different unitarization methods [45]; Arndt, Workman, Li and Roper's energy-independent partial wave analysis [48]; Nozawa, Blankleider, and Lee's calculation invoking off shell modeling of $\pi N$ interactions [47]; and a similar approach by Tanabe and Ohta where additional parameters are optimized by a chi-square fit [3].

Tanabe and Ohta's result for the $M1$ transition moment is relatively small compared to the other calculations reported in figure 14. They note however, that they have calculated the bare coupling and, as such, their result should be compared to the MIT bag model (the lower of the two entries in the Bag column of figure 14), as opposed to the Chiral bag model for example. They argue that their model explicitly takes into account the pion cloud effect separately. Similar arguments hold for the calculation of Nozawa, Blankleider, and Lee.

In this lattice calculation the $\Delta$ is stable and as a result the problems associated with defining the properties of a baryon unstable to strong interactions do not arise. In the lattice simulations, three quarks are created and allowed to propagate along paths in space-time determined by the QCD Lagrangian before they are later annihilated. The possible paths the quarks can take include paths such as $Z$-graphs where a quark emits a gluon and scatters into a negative energy state etc. At intermediate times there are 3 quarks and any number of quark-antiquark pairs in the $\Delta$ wave function. Presumably, these quark wave functions may be described in a Fock-space expansion of meson-baryon intermediate states, including $\pi N$ intermediate states. Baryon-antibaryon pairs may also play a role. In essence, the $\Delta$ baryon simulated on the lattice is dressed. However, it is not completely dressed since some diagrams corresponding to pion dressings are not included in making the quenched approximation.

The quark model calculations include (from top down) Guiasu and Koniuk's calculation in which mesonic dressings of the nucleon are explicitly included [40]; Capstick's calculation in which configuration mixing in the baryon SU(6) wave functions is accounted for [39]; and a calculation based on the simple quark model of Darewych, Horbatsch, and Koniuk [19].

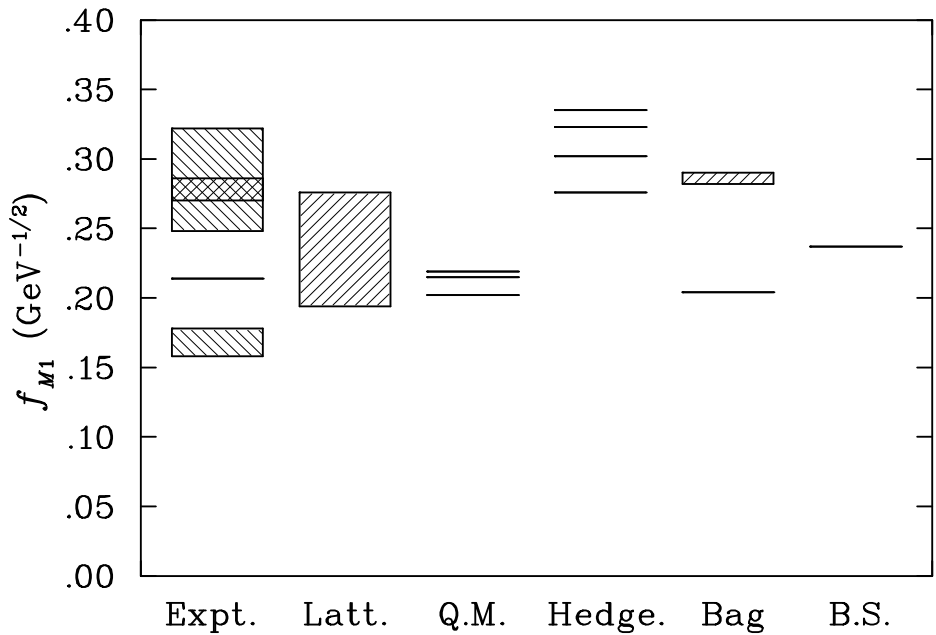

FIG. 14. Calculations of the magnetic transition moment, $f_{M1}$. The calculations have been divided among six categories including analyses of experimental pion photoproduction data (Expt.), our lattice QCD calculation (Latt.), nonrelativistic quark model determinations (Q.M.), hedgehog models (Hedge.), bag models (Bag), and a Bethe Salpeter calculation (B.S.).

It is often argued that the simple quark model does not include the physics of mesonic dressings. However we have seen remarkable agreement between the lattice calculations and simple constituent quark models in the decuplet baryon analysis and now in the transition moments under investigation here. It is important to ask what physics is represented by the constituent quark. In simple quark models, nonperturbative gluon interactions with current quarks (which naturally includes physics associated with quark-antiquark pairs and thus mesons) are approximated through the use of a constituent quark with an effective mass. In fact, the constituent quark masses are determined predominantly by reproducing nucleon properties such as the magnetic moment. Of course, proton properties reflect physics which may be ascribed to the associated pion cloud. Therefore constituent quark model predictions of magnetic moments and magnetic transition moments *implicitly* include the physics of mesonic dressings. As a result, it is not appropriate to directly compare the results of Tanabe and Ohta or Nozawa, Blankleider, and Lee with the quark model.

Guiasu and Koniuk have attempted to explicitly include the physics of the pion cloud in a quark model calculation of the helicity amplitudes of $N\,\gamma \leftrightarrow \Delta$ transitions. Of course, if one wishes to explicitly include pion dressings in the quark model, one must recalculate the constituent quark parameters. To do this, Guiasu and Koniuk recalculated the octet baryon magnetic moments in their new model. Perhaps it's not too surprising that their new result is largely unchanged from the simplest quark model. Their approach simply took some of the pion cloud physics implicitly contained in the constituent quarks and moved it to an explicit pion cloud.

Hedgehog models appear to predict values for $f_{M1}$ which are generally larger than the lattice prediction. The results presented here are obtained by taking the ratio of transition to proton moments and scaling the result such that the model calculations reproduce the proton moment. This approach eliminates, to some extent, the sensitivity of the hedgehog results to differences in the parameter sets of different authors. From top down, the calculations include the hybrid model of Cohen and Broniowski [49]; an early SU(2) Skyrme model calculation by Adkins, Nappi and Witten [50]; SU(2) Skyrme model calculations by Kunz and Mulders [51]; and an SU(3) Skyrme model calculation by Chemtob [52].

The bag models include an old MIT bag calculation of Donoghue, Golowich, and Holstein [53] and a chiral bag calculation of Kälbermann and Eisenberg [54]. The uncertainty region for the chiral bag calculation reflects the sensitivity of $f_{M1}$ on changing the bag radius from 0.8 to 1.0 fm. Finally the Bethe-Salpeter determination indicated in the final column of figure 14 is that of Mitra and Mittal [55]. A QCD sum rule result has not been included here because the approach is not able to provide a prediction for $f_{M1}$ that is free of unknown parameters [56].

Lattice predictions of the resonant contributions to the helicity amplitudes and radiative decay widths are summarized in table XI. The values are determined from the Sachs form factors using the relationships reviewed in the Appendix. Values for radiative branching ratios estimated using heavy baryon chiral perturbation theory [57] span a range which is consistent with our lattice predictions. Experimental estimates of the radiative decay widths are also given in table XI. The experimental limit for the $\Xi^{*0} \to \Xi^0 \gamma$ decay width is relatively close to our prediction. A nonvanishing experimental measurement of this decay width may be possible in the not too distant future.

TABLE XI. Lattice predictions for resonant contributions to the helicity amplitudes and radiative decay widths. Lattice results have been normalized to reproduce the proton magnetic moment. Experimental estimates of the radiative decay widths are also given.

| Transition | Lattice Predictions | | | Experiment |
| --- | --- | --- | --- | --- |
| | $A_{1/2}$ (GeV$^{-1/2}$) | $A_{3/2}$ (GeV$^{-1/2}$) | $\Gamma$ (MeV) | $\Gamma$ (MeV) |
| $p\,\gamma \leftrightarrow \Delta^+$ | $-0.125(41)$ | $-0.195(34)$ | $0.43(15)$ | $0.66(17)$[a] |
| $n\,\gamma \leftrightarrow \Delta^0$ | $0.125(41)$ | $0.195(34)$ | $0.43(15)$ | |
| $\Sigma^+\,\gamma \leftrightarrow \Sigma^{*+}$ | $-0.086(15)$ | $-0.125(20)$ | $0.100(26)$ | |
| $\Sigma^0\,\gamma \leftrightarrow \Sigma^{*0}$ | $-0.034(7)$ | $-0.052(8)$ | $0.017(4)$ | $<1.8$[b] |
| $\Sigma^-\,\gamma \leftrightarrow \Sigma^{*-}$ | $0.017(3)$ | $0.021(5)$ | $0.0033(12)$ | |
| $\Xi^0\,\gamma \leftrightarrow \Xi^{*0}$ | $0.082(11)$ | $0.130(15)$ | $0.129(29)$ | $<0.38$[b] |
| $\Xi^-\,\gamma \leftrightarrow \Xi^{*-}$ | $-0.016(2)$ | $-0.021(4)$ | $0.0038(12)$ | |

[a] Value calculated from the analysis of Davidson *et al.* [45].
[b] Values calculated from radiative branching ratios [58].

## VIII. SUMMARY AND OUTLOOK

We have presented a fully relativistic formalism for isolating and extracting the electromagnetic multipole form factors of spin-3/2 to spin-1/2 transitions in lattice field theory. Results of the first lattice QCD analysis of $SU(3)$-flavor octet to decuplet baryon transitions have been systematically examined to reveal new aspects of the underlying nonperturbative quark-gluon dynamics.

The $M1$ correlation functions for baryon transitions show statistical uncertainties similar to that seen in our octet baryon analysis. The central values were seen to drift to some extent in the plateau region for the light $u$- and $d$-quark three-point correlation functions. However, this is more likely an indication of the need for better statistics as opposed to a signature of the correlation functions failing to reach the plateau region before the lattice boundary is encountered.

Lattice calculations of the baryon octet to decuplet $M1$ transition moments agree with simple quark model predictions when the lattice results are scaled to reproduce the proton moment. Surprisingly, the manner in which the quarks contribute to the moments in the lattice calculations is quite different from that anticipated by the quark model. Ultimately, a high statistics calculation will reveal differences between the lattice and quark model results. However, differences in the results of the two approaches are small relative to the more dramatic effects seen in our octet baryon analysis. Here the effective quark moments indicate corrections to the simple quark model description of transition moments the order of 10 to 20%. In contrast, 50% corrections were seen in our octet baryon study. We look forward to an experimental determination of the hyperon $M1$ transition moments which will test these predictions.

Quenching of the effective quark moments in hyperons is seen in the $M1$ transitions. The quenching is larger in $\Xi$ than $\Sigma$. This is similar to our results for decuplet baryons. However, effective strange quark moments determined from transition moments are found to be suppressed relative to the values determined from decuplet baryon magnetic moments.

The lattice results prefer values for the ratio $\mu_{p\Delta}/\mu_p$ which are larger than simple quark model predictions. This is largely due to differences in the manner in which the quarks contribute to the proton moment.

The lattice prediction for the $M1$ transition moment of $N\,\gamma \leftrightarrow \Delta$ is in agreement with multipole analyses, nonrelativistic quark models, and a Bethe-Salpeter model approach, and suggests values smaller than that of the chiral bag model and typical values produced in hedgehog models. The statistical uncertainties in the lattice results for $\mathcal{R}_{EM}$ are relatively large. All the results of the experimental analyses lie comfortably within our prediction from first principles of $+(3 \pm 8)\%$ for $N\,\gamma \leftrightarrow \Delta$ transitions.

The $C2$ correlation functions become noisy at large time separations and a thorough examination of this form factor will have to wait for higher statistics calculations. The lattice results are consistent with expectations of multipole analyses.

The spatial asymmetry of our elongated lattice prevents us from drawing any strong conclusions on the $E2$ form factors. However, an important discovery is the manner in which statistical fluctuations are compensated in combining the quark contributions to negative charge hyperon transition moments. Calculations on a cubic lattice will provide precise estimates of the $E2$ transition moments for $\Sigma^-$ and $\Xi^-$ hyperons which will be indispensable in model development and testing.

A high statistics analysis of hadronic electromagnetic form factors would provide considerable insight to hadronic structure. Our present understandings of the essence of the underlying quark dynamics may be confirmed and refined. For example, the $E2$ moment of $\Delta$ is particularly interesting since it provides a glimpse into the shape of the baryon ground state. In general, hedgehog models such as the Skyrmion predict a large $E2$ moment. Our lattice results agree with hedgehog model predictions mainly due to the presence of large statistical uncertainties in the lattice predictions. The central value of the distribution suggests a smaller $E2$ moment. A high statistics lattice calculation would be able to confirm or reject the hedgehog Skyrmion description of baryons.

Statistically significant predictions for all the higher order multipole moments would be useful in the development of model hypotheses and evaluation of model predictions. A nonperturbative QCD determination of the $E2/M1$ ratio from first principles for the electromagnetic transition moments of $p\,\gamma \leftrightarrow \Delta^+$ with statistical uncertainties on par with the experimentally based determinations is anxiously awaited by those working in this field.

It is encouraging that *ab initio* lattice QCD calculations of electromagnetic multipole moments are already competitive with hadronic models which use adjustable parameters. With further investigations to reduce statistical and systematic errors, lattice studies of hadronic electromagnetic form factors will continue to provide new insight into nonperturbative QCD. The strong signals for magnetic dipole transitions for all octet baryons and for the electric quadrupole transitions of $\Sigma^-$ and $\Xi^-$ bode well for future lattice calculations of these quantities being able to further discriminate among models of hadronic structure.

## ACKNOWLEDGMENTS

The computing resources for this study were provided by the Computing Science Department and the Center for Computational Sciences at the University of Kentucky on their IBM 3090-600J supercomputer. D.B.L. thanks Satoshi Nozawa and Thomas Cohen for a number of helpful discussions. T.D. thanks Keh-Fei Liu for many useful conversations. This work is supported in part by the U.S. Department of Energy under grant numbers DE-FG05-87ER-40322 and DE-FG05-84ER-40154, the National Science Foundation under grant number STI-9108764 and the Natural Sciences and Engineering Research Council of Canada.

## APPENDIX: TRANSITION MOMENT PHENOMENOLOGY

In this section we make contact with other observables and formalisms associated with the phenomenology of electromagnetic transition moments. The following relationships are well established in the continuum. However, these relationships may not strictly hold for quenched lattice QCD. Since there are good reasons for calculating the Sachs form factors $(\mathcal{G}_{M1}, \mathcal{G}_{E2})$ in the lattice approach [44] as opposed to $f_{M1}$ and $f_{E2}$, we will use the continuum relationships with their parameters determined by experimental values. With this approach, the following relationships may be simply regarded as a form of "unit conversion".

The relationships between the Particle Data Group's [4] electromagnetic transition amplitudes $f_{M1}$ and $f_{E2}$ and the Sachs form factors investigated here are [47],

$$f_{M1} = \frac{e}{2M_N} \left( \frac{|\vec{q}| M_\Delta}{M_N} \right)^{1/2} \mathcal{G}_{M1}, \tag{A1a}$$

$$f_{E2} = -\frac{e}{2M_N} \left( \frac{|\vec{q}| M_\Delta}{M_N} \right)^{1/2} \frac{2|\vec{q}| M_\Delta}{M_\Delta^2 - M_N^2} \mathcal{G}_{E2}, \tag{A1b}$$

where $e = \sqrt{4\pi\alpha}$. In the rest frame of $\Delta$ at $q^2 = 0$, energy-momentum conservation sets $2|\vec{q}| M_\Delta = M_\Delta^2 - N_N^2$ and,

$$f_{M1} = \frac{e}{2M_N} \left( \frac{M_\Delta^2 - M_N^2}{2M_N} \right)^{1/2} \mathcal{G}_{M1}, \tag{A2a}$$

$$f_{E2} = -\frac{e}{2M_N} \left( \frac{M_\Delta^2 - M_N^2}{2M_N} \right)^{1/2} \mathcal{G}_{E2}. \tag{A2b}$$

The ratio of $E2$ to $M1$ form factors is defined by

$$\mathcal{R}_{EM} \equiv \frac{f_{E2}}{f_{M1}} = -\frac{\mathcal{G}_{E2}}{\mathcal{G}_{M1}}. \tag{A3}$$

Resonant contributions to the helicity amplitudes are given by simple linear combinations of $f_{M1}$ and $f_{E2}$:

$$A_{1/2} = -(f_{M1} + 3f_{E2})/2, \tag{A4a}$$

$$A_{3/2} = -\sqrt{3}(f_{M1} - f_{E2})/2. \tag{A4b}$$

Partial widths may also be inferred from the Sachs form factors assuming continuum dispersion relations:

$$\Gamma_{M1} = \frac{\alpha}{16} \frac{(M_\Delta^2 - M_N^2)^3}{M_N^2 M_\Delta^3} \mathcal{G}_{M1}^2, \tag{A5a}$$

$$\Gamma_{E2} = \frac{3\alpha}{16} \frac{(M_\Delta^2 - M_N^2)^3}{M_N^2 M_\Delta^3} \mathcal{G}_{E2}^2. \tag{A5b}$$